# Timelike fibers and four-dimensional mass: spectral bounds and a constrained bilocal oscillator in para-Kähler signature $(4,4)$

G. Kletetschka*[1,2]

[1]Faculty of Science, Charles University, Albertov 6, 128 43 Prague, Czech Republic

[2]Geophysical Institute, University of Alaska Fairbanks, Fairbanks, Alaska 99775, USA

*Corresponding author. E-mail: gkletetschka@alaska.edu; ORCID: 0000-0002-0645-9037

**Abstract**

We establish spectral bounds for the four-dimensional mass operator of a real scalar field with a two-derivative action on a warped product of Minkowski space with a pseudo-Riemannian fiber, and construct a constrained free model with positive four-dimensional masses. If the fiber metric has a timelike direction at an interior point, gradients along it enter the mass operator negatively and make it unbounded below for every warp factor, locally bounded potential and self-adjoint extension. On a negative-definite fiber, levels above the supremum of the warped potential require boundary terms. If a transitive isometry group of a connected negative-definite fiber without boundary preserves the warp factor and potential, an eigenvalue of finite multiplicity forces compactness and closed timelike geodesics. Flat para-Hermitian space of signature $(3,3)$ admits no Lorentz-algebra action by automorphisms. On flat para-Kähler space of signature $(4,4)$, whose points can be read as pairs of Minkowski points, the stabilizer of a Lorentzian Lagrangian section is the Lorentz group. For an input harmonic potential of frequency $\omega$ on the normal space, a bulk mass $m_\Phi$ with $m_\Phi^2/2\omega \notin \mathbb{Z}$ and a local Lorentz-covariant subsidiary condition, the free physical spectrum is $M_n^2 = m_\Phi^2 + 2\omega(n-1) > 0$, one scalar Poincaré representation per level, with positive norm and energy. The positive part is carried by the relative time, on which the kinetic metric is positive; the timelike normal directions contribute a negative zero-point shift. Preservation of the subsidiary condition is a further requirement for interactions. The local cubic vertex analyzed here violates it.

## 1 Introduction

The idea that the four-dimensional rest mass of a particle could be generated by its dynamics along additional temporal directions recurs in several forms. It is related to questions about the dimensionality of time [1, 2], to the instabilities of timelike Kaluza–Klein reduction [3–7], and to two-time physics, in which massive particles appear as gauge-fixed "shadows" of a system in signature $(d, 2)$ (Refs. [8–10]). It underlies proposals by the author in which particle masses are tied to structure along three temporal coordinates [11–13], later reformulated in para-Hermitian geometry [14, 15]. A companion manuscript, G. Kletetschka, "Lorentzian sections of para-Hermitian manifolds: a dimensional obstruction and a sharp transversality criterion" (submitted to J. Math. Phys., 2026; available from the author), analyzes the four-dimensional sections of this geometry. We cite it below as the companion paper, with the numbering of its theorems, corollaries, propositions, remarks, equations and appendixes in the version submitted on 31 October 2026. We pose the mathematical question:

> *Under which conditions can directions of timelike signature, transverse to a four-dimensional Lorentzian section, produce a discrete, positive, stable spectrum of four-dimensional masses?*

For the scalar operators of Assumption 1, the sign of the timelike gradient contribution is elementary. In the quadratic form of the mass operator, the gradient along a timelike direction enters with the sign of a tachyonic mass, so that it can only lower $M^2$ (Definition 2 and Remark 3); this is the mechanism behind the tachyonic towers of timelike Kaluza–Klein reduction. What requires proof is the scope of this statement and whether a construction outside it does better. We prove spectral bounds for warped fibers of arbitrary signature, potentials and self-adjoint extensions, and a homogeneity theorem showing that a finite-multiplicity mass level on a connected negative-definite fiber without boundary requires compactness when the isometry group acts transitively and preserves the data. Then we show that flat para-Hermitian space of signature $(3,3)$ admits no Lorentz-algebra action by automorphisms. Then we analyze a constrained free model on flat para-Kähler space of signature $(4,4)$, where Lorentz symmetry is realized exactly. In every case the timelike directions lower the mass, and every positive contribution we find comes from the potential, from boundary data, or from a direction on which the kinetic metric is positive.

**Main results.** Conventions are fixed in Sec. 2; $M^2 = -\eta(p,p)$ is the four-dimensional rest mass squared, so that $M^2 < 0$ is a tachyon.

*Theorem A* (sign identity and spectral bounds; Theorem 4, Corollary 5). Let a real scalar field with a two-derivative action live on a warped product $F \times_{e^\alpha} \mathbb{R}^{3,1}$ with fiber metric $\gamma$ and bulk mass function $W$ (Assumption 1). If $\gamma$ has a timelike direction at an interior point, the four-dimensional mass operator $L$ is unbounded below, for every warp factor, every locally bounded $W$ and every self-adjoint extension of $L|_{C_c^\infty(\mathrm{int}\,F)}$. If $\gamma$ is negative definite and no boundary terms arise, $\operatorname{spec} L$ lies below $\sup_F e^{2\alpha} W$. For $\alpha = 0$, $\gamma = -\delta$ on $F = \mathbb{R}^q$ and $W = m_\Phi^2 - \lambda^2 |y|^2$, the spectrum is the inverted tower $m_j^2 = m_\Phi^2 - \lambda(2j + q)$.

*Theorem B* (homogeneity trilemma; Theorem 7). Let $F$ be connected, of dimension $q \geq 1$ and without boundary, with $\gamma$ negative definite. At most two of the following hold: $F$ is non-compact; a transitive group of isometries of $F$ preserves $\alpha$ and $W$; $L$ has an eigenvalue of finite multiplicity. If the last two hold, $F$ is compact and the bulk carries a closed timelike geodesic through every point.

*Theorem C* (Lagrangian sections of flat $(4,4)$; Propositions 14 and 16, Corollary 17). On flat para-Kähler space $(\mathbb{R}^{4,4}, h, K)$, the stabilizer in $\mathrm{Aut}(h,K) \cong \mathrm{GL}(4,\mathbb{R})$ of the transverse section $\Sigma = \mathrm{graph}(\varphi)$ is $\{\Lambda : \Lambda^T \varphi \Lambda = \varphi\}$. If $h|_\Sigma$ is nondegenerate, the stabilizer is the full orthogonal group of $h|_\Sigma$ if and only if $\Sigma$ is Lagrangian, and then $\Sigma^\perp = K(\Sigma)$. For $\varphi = \eta$ it is $O(3,1)$, acting diagonally on coordinates $(z_1, z_2)$ in which $h = \eta(dz_1, dz_2)$; these are the doubled coordinates of double field theory. A free bulk field has $M^2 = m_\Phi^2 + \pi_{x_4}^2 - \sum_{i=1}^{3} \pi_{\tau_i}^2$, where $\partial_{x_4} = K\,\partial_{\tau_4}$ and the timelike normal directions $\tau_1, \tau_2, \tau_3$ are the relative spatial separation of the pair $(z_1, z_2)$.

*Theorem D* (a constrained bilocal oscillator; Theorem 19, Propositions 22, 24 and 25). Let $m_\Phi^2/(2\omega) \notin \mathbb{Z}$. With the potential $\omega^2 h(v,v)$ on the normal space and the local subsidiary condition of Definition 18, the physical one-particle states form one scalar unitary representation of the Poincaré group of $\Sigma$ for each level $M_n^2 = m_\Phi^2 + 2\omega(n-1) > 0$, with positive Klein–Gordon norm. The conserved canonical energy, unbounded above and below on general Cauchy data, is positive on the physical solutions, on which the timelike normal directions carry no dynamics. In the decomposition of Definition 2 the temporal contribution to every level is $-3\omega/2$, and the positive part of $M_n^2 - m_\Phi^2$ is carried by $x_4$. A local cubic vertex does not leave the physical subspace invariant, and the linearized Einstein–Hilbert sector without a compatibility condition contains three ghost vectors and three ghost moduli.

**What is standard and what is new.** Tachyonic Kaluza–Klein towers, matter instabilities and ghost vectors from timelike compactification are known [3–7], as is the ill-posedness of the Cauchy problem for ultrahyperbolic equations [2, 16, 17]. Theorem A is a Rayleigh-quotient argument, which we state in the generality needed here: warped fibers of arbitrary signature, locally bounded potentials and every self-adjoint extension. Theorem B combines it with a homogeneity argument; we have not found it in the literature. Propositions 11, 14 and 16 are elementary and fix the kinematics. Proposition 14 is the flat, Lorentzian instance of the relation between a para-Hermitian structure and a compatible generalized metric in Born geometry [18–21], in which spacetime is chosen as a Lagrangian submanifold [18]: $\Sigma$ and $K(\Sigma)$ are the eigenspaces of the generalized metric, and the stabilizer is the diagonal of the double Lorentz group of Ref. [21]. Proposition 16 identifies the space with the doubled space of double field theory [22–25] carrying the diagonal Lorentz action. The model of Theorem D is a bilocal oscillator in the tradition of Yukawa, Feynman–Kislinger–Ravndal, Kim and Noz, and Kim–Noz–Oh [26–30]. What is new is the combination of three features: the relative metric is $-\eta$, as forced by $h(K\,\cdot, K\,\cdot) = -h$; the subsidiary condition is local, because $K$ is the unique Lorentz-equivariant identification of $\Sigma$ with $N = K(\Sigma)$ up to scale; and we obtain the resulting spectrum and the positivity of the energy on the physical space. To our knowledge this sign and this condition have not been combined before. A physical identification of the tower requires further modeling and comparison with observations.

**Scope.** The absolute mass scale remains an input: the flat structure $(h, K)$ is dilation covariant, so the scale $\omega$ and the gap are inputs. Masses generated by interactions (bound states, condensates, confinement), spinor fields beyond the remarks in Sec. 9, and dynamical para-complex structures are outside the scope of this paper.

**Methods and verification.** All proofs are analytic and complete in the text and the appendixes. The only computer-assisted statements, on the cubic vertex for states in relative motion (after Proposition 24) and on the compatible metric sector (Sec. 7.3), are marked as such and are not used in any proof. The supplementary information (Online Resources 1 and 2) contains symbolic and numerical checks of the main statements, by methods independent of the analytic proofs, with their outputs; no proof depends on them. The author used the large language model Claude Opus 5.5 (Anthropic) to search the literature, to propose and check derivations, to write the verification scripts and to draft the text, and ChatGPT (OpenAI) for editorial revisions and to clarify the functional-analytic scope of the statements. The author checked every proof, script output, reference and suggested change and takes full responsibility for the content.

**Organization.** Section 2 clarifies conventions and the setting, and states what would count as mass generated by extra times. Section 3 proves the spectral bounds and the homogeneity theorem. Section 4 treats flat para-Hermitian space of signature (3,3), Sec. 5 the kinematics of signature (4,4), Sec. 6 the constrained oscillator and its Cauchy problem, and Sec. 7 its interactions and its metric sector. Section 8 relates the results to earlier proposals, and Sec. 9 summarizes what is established and what is open. Appendix A computes the interaction overlaps and Appendix B collects the Kaluza–Klein formulas.

# 2 Setting and criteria

## 2.1 Conventions

We set $\hbar = c = 1$ and use the mostly-plus convention. On the section $\Sigma \cong \mathbb{R}^{3,1}$ we write $\eta = \mathrm{diag}(1,1,1,-1)$ in coordinates $(x^1, x^2, x^3, x^0)$, $\Box = \eta^{\mu\nu}\,\partial_\mu\,\partial_\nu$, and $M^2 = -\eta(p,p)$, so that $\Box e^{ip\cdot x} = M^2 e^{ip\cdot x}$. A vector $v$ is timelike for a metric $g$ if $g(v,v) < 0$. The bulk mass of a scalar field is $m_\Phi$; when a potential $V$ is present we write $W = m_\Phi^2 + V$.

The flat para-Hermitian model on $\mathbb{R}^{n,n}$ of Ref. [14] and of the companion paper has coordinates $x_1, \dots, x_n$ and $\tau_1, \dots, \tau_n$, with $\tau_k = ct_k$, and

$$h = \sum_{k=1}^{n} \left(dx_k^2 - d\tau_k^2\right), \qquad K\,\partial_{x_k} = \partial_{\tau_k}, \quad K\,\partial_{\tau_k} = \partial_{x_k}, \tag{1}$$

so that $K^2 = \mathbb{1}$, $h(K\,\cdot, K\,\cdot) = -h$, and the fundamental form $\Omega = h(K\,\cdot,\cdot)$ is closed; $(h, K)$ is para-Kähler. The eigenspaces $V_\pm = \ker(K \mp \mathbb{1})$ are spanned by $\partial_{u_k}$ and $\partial_{v_k}$, with $u_k = x_k + \tau_k$ and $v_k = x_k - \tau_k$; in these coordinates $h = \sum_k d\,u_k\,dv_k$, each eigenspace is totally $h$-null, and $h\left(\partial_{u_k}, \partial_{v_l}\right) = 1/2\,\delta_{kl}$. A linear map commuting with $K$ preserves $V_+$ and $V_-$, and it preserves $h$ exactly when its block on $V_-$ is the inverse transpose of its block on $V_+$ with respect to this pairing; hence the group of linear maps preserving $h$ and $K$ is

$$\mathrm{Aut}(h, K) \cong \mathrm{GL}(n, \mathbb{R}), \qquad u \mapsto \Lambda u, \quad v \mapsto \Lambda^{-T} v, \tag{2}$$

the intersection $O(n, n) \cap Sp(2n, \mathbb{R})$ familiar from Born geometry [20]. A section is a linear subspace $\Sigma$; it is transverse if $\Sigma \cap V_\pm = 0$. If $\Sigma$ is transverse, the projection of $\Sigma$ onto $V_+$ along $V_-$ is injective; hence a four-dimensional transverse section requires $n \geq 4$, and for $n = 4$ it is the graph $\Sigma = \{u + \varphi u : u \in V_+\}$ of an invertible linear map $\varphi: V_+ \to V_-$. This is the linear case of the dimensional obstruction of the companion paper (Theorem 10 and Corollary 11 there), which also analyzes the four-dimensional sections of the case $n = 3$.

## 2.2 Setting

**Assumption 1.**

(H1) The bulk is a warped product $\mathcal{M} = F \times_{e^\alpha} \mathbb{R}^{3,1}$, in the notation of Ref. [31], with metric

$$g = e^{2\alpha(y)} \eta_{\mu\nu} dx^\mu dx^\nu + \gamma_{ab}(y)\, dy^a dy^b, \tag{3}$$

where $(F, \gamma)$, the fiber, is a connected $q$-dimensional pseudo-Riemannian manifold, possibly with boundary, and $\alpha$ is smooth.

(H2) The field is a real scalar with action

$$S = \int_{\mathcal{M}} d^4 x\, d^q y \sqrt{|g|}\left[-1/2\, g^{AB}\, \partial_A \Phi\, \partial_B \Phi - 1/2\, W(y) \Phi^2\right], \tag{4}$$

with $W$ real and locally bounded.

(H3) The fiber operator $L$ of (5) below is realized as a self-adjoint operator in the complex Hilbert space $\mathcal{H} = L^2(F, \varrho\, e^{-2\alpha} d^q y)$ that extends $L|_{C_c^\infty(\mathrm{int}\,F)}$, where $\varrho = \sqrt{|g|} = e^{4\alpha}\sqrt{|\gamma|}$.

Separating $\Phi = \phi(x)\chi(y)$ in the field equation $g^{AB} \nabla_A \nabla_B \Phi = W\Phi$ gives $\square\phi = M^2 \phi$ with

$$L\chi \equiv e^{2\alpha}\left[W\chi - \varrho^{-1}\, \partial_a\left(\varrho\, \gamma^{ab}\, \partial_b \chi\right)\right] = M^2 \chi. \tag{5}$$

The operator $L$ is symmetric on $C_c^\infty(\mathrm{int}\,F) \subset \mathcal{H}$, and the four-dimensional mass spectrum is $\mathrm{spec}\,L$. Since $L$ commutes with complex conjugation, its deficiency indices are equal and self-adjoint extensions exist (von Neumann's theorem [32], Sec. X.1), so (H3) can always be met; the extension is in general not unique, for instance on a fiber with boundary (as for the Robin conditions of Remark 6), and the results below hold for each of them.

## 2.3 What would count as mass generated by extra time dimensions

We assess a direct temporal origin of mass by the following criteria. Here temporal origin refers specifically to the additive contribution of timelike gradients to the quadratic form of the mass operator, in the sense of Definition 2; a positive physical spectrum selected by constraints, as in Secs. 5–7, is assessed against the same criteria in Sec. 7.1. A sector realizes a direct temporal origin of mass if it satisfies:

(P1) *Positivity:* $\inf \operatorname{spec} M^2 \geq m_0^2 > 0$ on the physical state space.

(P2) *Temporal origin:* the levels are discrete and of finite multiplicity, and $M^2 - m_\Phi^2$ contains a positive contribution generated by the extra temporal directions in the sense of Definition 2.

(P3) *Norm:* the physical inner product is positive, including the sector of metric fluctuations.

(P4) *Evolution:* the dynamics is well posed in the four-dimensional time, with stated Cauchy data on a hypersurface $\{x^0 = t\} \times F$.

(P5) *No closed timelike geodesics:* the extra timelike directions are non-compact and there are no closed timelike geodesics.

(P6) *Symmetry:* translations along $\Sigma$ are exact; translations along the extra directions are exact or gauge, or they are broken explicitly by a function of the extra coordinates, whose contributions to $M^2$ then count as inputs.

(P7) *Lorentz invariance:* exact, and realized by automorphisms of the bulk structure.

Every bulk with at least two timelike directions, in particular every bulk considered in this paper, violates the chronology condition: a circle in a negative-definite 2-plane, for instance the plane of $x^0$ and a timelike fiber direction, is a closed timelike curve. Criterion (P5) asks only that freely falling particles do not return to their past.

**Definition 2** (temporal contribution)**.** Let $TF = T_+ \oplus T_-$ be a smooth $\gamma$-orthogonal splitting with $\gamma$ positive definite on $T_+$ and negative definite on $T_-$, and write $\gamma^{-1} = \gamma_+^{-1} + \gamma_-^{-1}$ for the induced splitting of the inverse metric, with $\gamma_+^{-1}$ positive and $\gamma_-^{-1}$ negative semidefinite. For $\chi$ in the domain of $L$ with $\|\chi\| = 1$ for which the integrals converge, put

$$\begin{aligned} \mathcal{W}[\chi] &= \int_F \varrho\, W|\chi|^2, \\ \mathcal{S}[\chi] &= \int_F \varrho\, \gamma_+^{ab}\, \partial_a\bar{\chi}\, \partial_b\chi \geq 0, \\ \mathcal{T}[\chi] &= \int_F \varrho\, \gamma_-^{ab}\, \partial_a\bar{\chi}\, \partial_b\chi \leq 0, \end{aligned} \tag{6}$$

with $d^q y$ understood, and $\mathcal{B}[\chi] = \langle \chi, L\chi \rangle - \mathcal{W}[\chi] - \mathcal{S}[\chi] - \mathcal{T}[\chi]$. We call $\mathcal{T}[\chi]$ the temporal contribution, $\mathcal{S}[\chi]$ the spacelike gradient contribution, and $\mathcal{W}[\chi]$ and $\mathcal{B}[\chi]$ the contributions of the inputs, the bulk mass function and the boundary data. A level $M^2$ contains a positive contribution generated by the extra temporal directions if $\mathcal{T}[\chi] > 0$ for a normalized eigenfunction, or along a normalized Weyl sequence, of that level and for some such splitting.

*Remark 3.* Since $\gamma_-^{-1}$ is negative semidefinite, $\mathcal{T}[\chi] \leq 0$ for every $\chi$ and every splitting. Hence no level of an operator of the form (5), with or without a subsidiary condition, contains a positive contribution generated by the extra temporal directions: this sign alone excludes a direct temporal origin of mass, both under Assumption 1 and after any restriction of the state space. This is the elementary content of the tachyonic towers of timelike Kaluza–Klein reduction [4, 5]. The results below establish its scope. By Theorem 4(b), (P1) fails on the full state space as soon as $\gamma$ has a timelike direction, so a sector satisfying (P1) requires a restriction of the state space, which lies outside Assumption 1. By Theorem 4(c), on a negative-definite fiber a level above $\sup_F e^{2\alpha} W$ requires a positive boundary term. By Theorem 7, discreteness together with homogeneity requires compact timelike directions. Sections 5–7 analyze a restriction by a subsidiary condition in signature (4,4).

# 3 Spectral bounds and homogeneity

## 3.1 The sign identity

**Theorem 4** (sign identity and spectral bounds). *Under Assumption 1:*

*(a)* *For every* $\chi \in C_c^\infty(\mathit{int}\, F)$,

$$\langle \chi, L\chi \rangle = \int_F \varrho \left(W|\chi|^2 + \gamma^{ab}\, \partial_a \bar{\chi}\, \partial_b \chi\right) d^q y, \qquad \|\chi\|^2 = \int_F \varrho\, e^{-2\alpha} |\chi|^2\, d^q y, \tag{7}$$

*that is,* $\mathcal{B}[\chi] = 0$ *in Definition 2.*

*(b)* *If* $\gamma$ *has a timelike direction at some interior point, then* $\mathit{inf\ spec}\, L = -\infty$ *for every self-adjoint extension of* $L|_{C_c^\infty(\mathit{int}\, F)}$.

*(c)* *If* $\gamma$ *is negative definite,* $\mathit{sup}_F\, e^{2\alpha} W < \infty$, *and the bound* $\langle \chi, L\chi \rangle \le \int_F \varrho\ W|\chi|^2 d^q y$, *which holds on* $C_c^\infty(\mathit{int}\, F)$ *by (a), holds on the domain of the extension (no positive boundary terms; for instance for Dirichlet or Neumann conditions on a compact F with smooth boundary, or when L is essentially self-adjoint on* $C_c^\infty(\mathit{int}\, F)$*), then* $\mathit{sup\ spec}\, L \le \mathit{sup}_F\, e^{2\alpha} W$.

*Proof.* (a) Multiply (5) by $\varrho e^{-2\alpha} \bar{\chi}$ and integrate by parts; there are no boundary terms for compact support in the interior.

(b) At an interior point $y_0$ choose a neighborhood $U$, a smooth real function $\xi$ with $\gamma^{ab}\, \partial_a \xi\, \partial_b \xi \le -c < 0$ on $U$, and a real $f \in C_c^\infty(U)$, $f \ne 0$. For $\chi_k = f \cos(k\xi) \in C_c^\infty(U)$,

$$\gamma^{ab}\, \partial_a \chi_k\, \partial_b \chi_k = k^2 f^2 \sin^2(k\xi)\, \gamma^{ab}\, \partial_a \xi\, \partial_b \xi + O(k) \le -ck^2 f^2 \sin^2(k\xi) + O(k),$$

uniformly on $U$. Since $d\xi \ne 0$ on $U$, $\int_U \varrho f^2 \sin^2(k\xi) \to 1/2 \int_U \varrho f^2$ and $\|\chi_k\|^2 \to 1/2 \int_U \varrho\, e^{-2\alpha} f^2$, while $\int \varrho W \chi_k^2$ stays bounded. Hence

$$\frac{\langle \chi_k, L\chi_k \rangle}{\|\chi_k\|^2} \le -c' k^2 + O(k) \longrightarrow -\infty.$$

For a self-adjoint operator $T$ and $\chi$ in its domain, $\langle \chi, T\chi \rangle \ge (\inf \operatorname{spec} T) \|\chi\|^2$ by the spectral theorem [33]; the domain contains $C_c^\infty(\operatorname{int} F)$ by (H3). Therefore $\inf \operatorname{spec} L = -\infty$.

(c) If $\gamma$ is negative definite, $\langle \chi, L\chi \rangle \le \int \varrho W |\chi|^2 \le \sup(e^{2\alpha} W)\, \|\chi\|^2$ for $\chi$ in the domain, and $\sup \operatorname{spec} T = \sup_{\chi \ne 0} \langle \chi, T\chi \rangle / \|\chi\|^2$ for self-adjoint $T$ (Ref. [33]). If $L$ is essentially self-adjoint on $C_c^\infty(\operatorname{int} F)$, the bound passes from $C_c^\infty(\operatorname{int} F)$ to the domain of the closure by approximation in the graph norm. □

Part (c) says that on a negative-definite fiber, gradients along the extra times raise the four-dimensional mass of no mode above the largest value of the warped bulk mass function; whatever a mode carries beyond that is supplied by boundary terms. The sign in part (a) is the origin of the tachyonic Kaluza–Klein towers of timelike compactification [4, 5]. Brane worlds with timelike extra dimensions appear in Refs. [5, 34].

**Corollary 5** (towers). *Under Assumption 1 with* $\alpha = 0$ *and flat negative-definite* $\gamma = -\delta$*:*

*(i)* *On* $F = \mathbb{R}^q$ *with* $W = m_\Phi^2$, $L = m_\Phi^2 + \Delta_y$ *is essentially self-adjoint on* $C_c^\infty(\mathbb{R}^q)$, *and* $\mathit{spec}\, L = (-\infty, m_\Phi^2]$, *purely continuous.*

*(ii)* *On the torus* $(\mathbb{R}/2\pi R\mathbb{Z})^q$ *with* $W = m_\Phi^2$, $L$ *is essentially self-adjoint on* $C^\infty$, *with eigenvalues* $m_{\boldsymbol{n}}^2 = m_\Phi^2 - |\boldsymbol{n}|^2/R^2$, $\boldsymbol{n} \in \mathbb{Z}^q$, *unbounded below.*

*(iii)* *With* $W = m_\Phi^2 - \lambda^2|y|^2$ *on* $F = \mathbb{R}^q$, $\lambda > 0$, $L = m_\Phi^2 - \left(-\Delta_y + \lambda^2|y|^2\right)$ *is essentially self-adjoint on* $C_c^\infty(\mathbb{R}^q)$, *and*

$$\operatorname{spec} L = \left\{m_j^2 = m_\Phi^2 - \lambda(2j + q)\colon\ j = 0,1,2,\dots\right\}, \tag{8}$$

*with the multiplicity* $\binom{j+q-1}{q-1}$ *of the* $q$*-dimensional isotropic oscillator. For* $q = 2$, $m_j^2 = m_\Phi^2 - 2\lambda(j + 1)$ *with multiplicity* $j + 1$.

*(iv)* *With* $W = m_\Phi^2 + \lambda^2|y|^2$ *on* $F = \mathbb{R}^q$, $L = m_\Phi^2 - \left(-\Delta_y - \lambda^2|y|^2\right)$ *is essentially self-adjoint on* $C_c^\infty(\mathbb{R}^q)$, *and its spectrum is* $\mathbb{R}$, *purely continuous.*

*Proof.* Cases (i) and (ii) follow by Fourier analysis. In (iii), $-\Delta + \lambda^2|y|^2$ is essentially self-adjoint on $C_c^\infty(\mathbb{R}^q)$ by Kato's theorem [32], and the functions $\lambda^{q/4} h_{n_1}\left(\sqrt{\lambda}\, y_1\right) \cdots h_{n_q}\left(\sqrt{\lambda}\, y_q\right)$, $n \in \{0,1,2,\dots\}^q$, with $h_k$ the Hermite functions, form an orthonormal basis of $L^2(\mathbb{R}^q)$ of eigenfunctions with eigenvalues $\lambda(2|n| + q)$, $|n| = n_1 + \cdots + n_q$; since exactly $\binom{j+q-1}{q-1}$ multi-indices have $|n| = j$, this gives (8) and the multiplicities. In (iv), essential self-adjointness follows from the Faris–Lavine theorem [32], since the potential $-\lambda^2|y|^2$ is bounded below by a multiple of $-|y|^2$. In one dimension put $a = (p + \lambda y)/\sqrt{2\lambda}$ and $b = (p - \lambda y)/\sqrt{2\lambda}$, with $p = -i\,\partial_y$; then $[a, b] = i$ and $p^2 - \lambda^2 y^2 = \lambda(ab + ba)$. By the Stone–von Neumann theorem the pair $(a, b)$ is unitarily equivalent to $(y, p)$, so $-\partial_y^2 - \lambda^2 y^2$ is unitarily equivalent to $\lambda(yp + py)$, twice $\lambda$ times the generator of dilations, whose spectrum is $\mathbb{R}$ and purely continuous. The $q$-dimensional operator is a sum of $q$ commuting copies. □

*Remark 6.* Case (iii) is the only discrete tower on a non-compact fiber in Corollary 5, and it requires $W \to -\infty$ at large $|y|$: for this harmonic potential on a flat timelike fiber, the bulk mass function is tachyonic far from the origin and the resulting tower is inverted and unbounded below. Case (iv) shows that a bulk mass function growing at infinity does not confine. A Robin condition $\chi'(0) = -\kappa\chi(0)$, $\kappa > 0$, on a timelike half-line binds an isolated level at $m_\Phi^2 + \kappa^2$, above $m_\Phi^2$, with eigenfunction $\sqrt{2\kappa}\, e^{-\kappa y}$; in the decomposition (6) it has $\mathcal{W} = m_\Phi^2$, $\mathcal{T} = -\kappa^2$ and $\mathcal{B} = 2\kappa^2$, so the level is supplied by the boundary datum $\kappa$, consistently with Theorem 4(c). The essential spectrum $(-\infty, m_\Phi^2]$ is that of the Dirichlet realization, by Weyl's theorem [35], since the two resolvents differ by an operator of rank one. For a charged complex field, which lies outside Assumption 1, a uniform magnetic field $B$ through a timelike 2-plane gives the inverted Landau ladder $m_\Phi^2 - (2n + 1)|eB|$, $n = 0,1,2,\dots$, each level infinitely degenerate (Online Resource 1, Sec. S2.2).

## 3.2 The homogeneity trilemma

Corollary 5 exhibits two ways to obtain discrete levels: compactness, or an explicit function of the extra times. The next theorem gives the corresponding obstruction when the data are homogeneous: if a transitive group of isometries of a connected negative-definite fiber without boundary preserves the warp factor and the potential, an eigenvalue of finite multiplicity forces the fiber to be compact.

**Theorem 7** (homogeneity trilemma)**.** *Let Assumption 1 hold with* $F$ *connected, of dimension* $q \geq 1$ *and without boundary, and* $\gamma$ *negative definite. Suppose a group* $G$ *acts transitively on* $F$ *by isometries of* $\gamma$ *that preserve* $\alpha$ *and* $W$*. If* $L$ *has an eigenvalue of finite multiplicity, then* $F$ *is compact and* $\mathcal{M}$ *contains a closed timelike geodesic through every point. Hence at most two of the following hold: (i)* $F$ *is non-compact; (ii) such a group* $G$ *exists; (iii)* $L$ *has an eigenvalue of finite multiplicity.*

*Proof.* A $G$-invariant function on a homogeneous space is constant, so $\alpha$ and $W$ are constant; then $\mathcal{M}$ is a metric product, and after rescaling $x$ we may take $\alpha = 0$, so that $L = W + \Delta_F$, with $\Delta_F$ the Laplace–Beltrami operator of the Riemannian metric $-\gamma$, and $\mathcal{H} = L^2(F, d\,\mathrm{vol})$ up to a constant factor. A homogeneous Riemannian manifold is complete: if the exponential map at one point is defined on the ball of radius $\epsilon$, then by homogeneity it is so at every point, every geodesic can be continued, and the Hopf–Rinow theorem applies [36]. On a complete Riemannian manifold the Laplace–Beltrami operator is essentially self-adjoint on $C_c^\infty(F)$ (Ref. [37]); hence the realization of (H3) is unique and commutes with the action $(U_g f)(y) = f(g^{-1}y)$ of $G$, which is unitary because isometries preserve $d\,\mathrm{vol}$. Let $\mathcal{E}$ be an eigenspace of finite dimension $d$, with orthonormal basis $f_1, \dots, f_d$; the $f_i$ are smooth by elliptic regularity. The function $S = \sum_i |f_i|^2$ is independent of the choice of orthonormal basis; $G$ preserves $\mathcal{E}$ and acts on it unitarily, so $S(g^{-1}y) = S(y)$ for all $g \in G$. (This is the argument behind the addition theorem for spherical harmonics.) By transitivity $S$ is a constant $s$, and $s\,\mathrm{vol}(F) = \int_F S = d$. Hence $\mathrm{vol}(F) < \infty$.

Isometries map metric balls onto metric balls, so all balls of a fixed radius $r$ have the same positive volume. If $\mathrm{vol}(F) < \infty$, a maximal family of disjoint balls of radius $r/2$ is finite, the balls of radius $r$ with the same centers cover $F$, and since $F$ is connected it is bounded, hence compact by the Hopf–Rinow theorem [36]. A compact Riemannian manifold of positive dimension carries a non-trivial closed geodesic [38, 39]. In the product $\mathcal{M}$ the fibers $F \times \{x\}$ are totally geodesic, and a non-constant curve in them is timelike because $\gamma$ is negative definite. Translating the geodesic by $G$ and along $\mathbb{R}^{3,1}$ gives one through every point. □

Theorem 7 makes clear why discrete masses from homogeneous extra times require compact timelike directions. Escapes outside its hypotheses, such as breaking the translations along the extra times by a potential (Corollary 5(iii)) or gauging them, are not covered by it.

### 3.3 Metric fluctuations and evolution

**Proposition 8** (graviphoton signs)**.** *In the Kaluza–Klein reduction of the Einstein–Hilbert action on the product (3) with $\alpha = 0$ and constant fiber metric $\gamma_{ab}$, with the overall sign of the action fixed by requiring a healthy four-dimensional graviton, the graviphotons $A^a_\mu$ of the fiber-independent sector have kinetic term $-1/4\,\gamma_{ab}F^a_{\mu\nu}F^{b\,\mu\nu}$ per unit fiber volume. Each timelike direction of $\gamma$ contributes a vector field whose kinetic term has the wrong sign, a ghost.*

*Proof.* This is the standard Kaluza–Klein computation [40], Eqs. (5), (9) and (31), with a constant fiber metric of arbitrary signature; the formulas are collected in Appendix B. In the mostly-plus convention a vector field is healthy when its kinetic term is $-1/4\,F^2$. For non-compact $F$ the fiber-independent modes are not normalizable, and the statement refers to the consistent truncation to these modes, per unit fiber volume. □

For a single timelike circle the sign reversal of the graviphoton kinetic term is stated in Ref. [6], Sec. 4, and for the negative-definite fiber of the $(3,3)$ model of Ref. [14] in the companion paper, Appendix A. The ghosts of Proposition 8 are vector fields; they are not the scalar from the time component of a gauge field, which can be removed by a gauge choice [6, 7].

**Proposition 9** (ill-posedness of the ultrahyperbolic Cauchy problem)**.** *Let $s \geq 0$ and $t \geq 2$. For the flat equation $(\Delta_x - \Delta_\tau)\Phi = m^2\Phi$ on $\mathbb{R}^{s,t}$, with data $(\Phi, \partial_{\tau_1}\Phi)$ on $\{\tau_1 = 0\}$, there is no continuous solution map from $H^r \times H^{r'}$ to $H^{r''}(\{\tau_1 = T\})$, for any $T \neq 0$ and any $r, r', r'' \in \mathbb{R}$.*

*Proof.* Write $\tau' = (\tau_2, \dots, \tau_t)$ and let $(\mathbf{k}, \boldsymbol{\ell})$ be the dual variables of $(x, \tau')$. On the set where $|\boldsymbol{\ell}|^2 > |\mathbf{k}|^2 + m^2$, the function $\Gamma = (|\boldsymbol{\ell}|^2 - |\mathbf{k}|^2 - m^2)^{1/2}$ is the growth rate of the solutions $e^{i\mathbf{k}\cdot x + i\boldsymbol{\ell}\cdot\tau'}e^{\pm\Gamma\tau_1}$. Take data $\hat{\Phi}_0 = \hat{f}_N$ and $\partial_{\tau_1}\hat{\Phi}_0 = \pm\Gamma\hat{f}_N$, with $\hat{f}_N$ a fixed bump function translated to the frequency $(\mathbf{0}, \boldsymbol{\ell}_N)$, $|\boldsymbol{\ell}_N| = N$, the sign chosen as that of $T$. The solution has $\hat{\Phi}(T) = e^{\Gamma|T|}\hat{f}_N$. On the support of $\hat{f}_N$, $\Gamma \geq N/2$ for large $N$, so $\|\Phi(T)\|_{H^{r''}} \geq c\, e^{N|T|/2}(1+N)^{-|r''|}$, while the data norms grow at most polynomially in $N$. Hence no estimate of the form $\|\Phi(T)\|_{H^{r''}} \leq C(\|\Phi_0\|_{H^r} + \|\partial_{\tau_1}\Phi_0\|_{H^{r'}})$ holds. □

The ill-posedness is classical [2, 16]. Craig and Weinstein [17] treat the massless case, show that well-posedness is restored by nonlocal constraints on the data, and analyze surfaces of higher codimension; the mass term does not change the argument. This is the standard objection to evolution in several time dimensions. The same mechanism operates for evolution in the time of the section once the normal space contains timelike directions: in the model of Sec. 6, the unconstrained Cauchy problem has modes with unbounded growth rates (Proposition 22).

# 4 Flat para-Hermitian space of signature $(3,3)$

**Lemma 10** (normal bundle)**.** *Let $U \subset \mathbb{R}^{s,t}$ be a nondegenerate subspace of signature* $(3,1)$*. Then $U^\perp$ has signature $(s-3, t-1)$ and is the orthogonal sum of $\min(s-3, t-1)$ hyperbolic planes and a definite subspace of dimension $|s-t-2|$. For $(s,t) = (3,3)$, $U^\perp$ is negative definite of dimension two; for $(4,4)$ it has signature $(1,3)$.*

*Proof.* Sylvester's law of inertia and the Witt decomposition [41]. □

In signature $(3,3)$ both normal directions of a section of signature $(3,1)$ are timelike, and Theorem 4 applies with a negative-definite fiber.

**Proposition 11.** *Every homomorphism of Lie algebras $\mathfrak{so}(3,1) \to \mathfrak{gl}(3,\mathbb{R})$ is zero. Consequently every homomorphism from $\mathfrak{so}(3,1)$ to the Lie algebra $\mathfrak{gl}(3,\mathbb{R}) \ltimes \mathbb{R}^6$ of affine automorphisms of the flat structure $(h,K)$ on $\mathbb{R}^{3,3}$ is zero, and no subgroup of $Aut(h,K) \cong GL(3,\mathbb{R})$ is locally isomorphic to $SO(3,1)$.*

*Proof.* Let $J_i$ and $B_i$ be rotation and boost generators, with $[J_i, J_j] = \epsilon_{ijk}J_k$, $[J_i, B_j] = \epsilon_{ijk}B_k$ and $[B_i, B_j] = -\epsilon_{ijk}J_k$, and let $\rho\colon \mathfrak{so}(3,1) \to \mathfrak{gl}(3,\mathbb{R})$ be a homomorphism. Its restriction to $\mathfrak{so}(3)$ is completely reducible, since $\mathfrak{so}(3)$ is semisimple. The complexification of a real irreducible representation of $\mathfrak{so}(3)$ is either irreducible, of odd dimension $2j+1$, or the sum of two copies of an irreducible representation of half-integer spin $j$, of dimension $2(2j+1) \geq 4$; hence $\mathbb{R}^3$ is either the trivial representation or the vector representation of $\mathfrak{so}(3)$. If it is trivial, $\rho(B_k) = 1/2\, \epsilon_{ijk}[\rho(J_i), \rho(B_j)] = 0$, so $\rho = 0$. Otherwise we may take $\rho(J_i) = J_i$, the generators of $SO(3)$. The relation $[J_i, B_j] = \epsilon_{ijk}B_k$ says that $(\rho(B_i))$ is an $\mathfrak{so}(3)$-equivariant map from the vector representation to $\mathfrak{gl}(3,\mathbb{R}) = \mathbb{R} \oplus \mathfrak{so}(3) \oplus \mathrm{Sym}^2_0$ (dimensions $1+3+5$). The vector representation occurs once and is absolutely irreducible, so by Schur's lemma $\rho(B_i) = c\,J_i$ for some $c \in \mathbb{R}$. Then $[\rho(B_i), \rho(B_j)] = c^2\epsilon_{ijk}J_k$, while the boost relation requires $-\epsilon_{ijk}J_k$; hence $c^2 = -1$, which has no real solution.

Isometries of the flat metric $h$ are affine, and an affine map preserves $K$ if and only if its linear part commutes with $K$; by (2) the Lie algebra of affine automorphisms is $\mathfrak{gl}(3,\mathbb{R}) \ltimes \mathbb{R}^6$, with $\mathbb{R}^6$ the abelian ideal of translations. A homomorphism $\rho$ from $\mathfrak{so}(3,1)$ into it projects to zero in $\mathfrak{gl}(3,\mathbb{R})$, so it takes values in $\mathbb{R}^6$, and $\rho([X,Y]) = [\rho(X), \rho(Y)] = 0$; since $\mathfrak{so}(3,1)$ is perfect, $\rho = 0$. □

The proof uses only the sign that distinguishes $\mathfrak{so}(3,1)$ from $\mathfrak{so}(4)$; for $\mathfrak{so}(4) = \mathfrak{so}(3) \oplus \mathfrak{so}(3)$ the choice $c = \pm 1$ gives the two three-dimensional representations.

**Proposition 12** (gauge reductions)**.** *For a relativistic particle in $\mathbb{R}^{s,t}$ with phase-space coordinates $(X,P)$ and mass-shell constraint $P \cdot P + m^2 = 0$:*

(i) *the* $\mathfrak{sp}(2,\mathbb{R})$ *constraints* $X\cdot X = X\cdot P = P\cdot P = 0$, *which replace the mass shell, give a reduced phase space symplectomorphic to that of a massless particle in signature* $(s-1,t-1)$ *(Refs. [8, 9]);*

(ii) *for a null vector* $\ell$ *and* $c\in\mathbb{R}$, *the first-class constraints* $X\cdot\ell = c$, $P\cdot\ell = 0$ *reduce it to a particle of the same mass in* $\ell^\perp/\mathbb{R}\ell$, *of signature* $(s-1,t-1)$*;*

(iii) *for a unit vector* $w$ *with* $w\cdot w=\epsilon=\pm1$, *the constraint* $P\cdot w=\beta$ *with the gauge condition* $X\cdot w=0$ *reduces it to a particle in* $w^\perp$ *with mass squared* $m^2+\epsilon\beta^2$, *of signature* $(s,t-1)$ *for timelike* $w$ *and* $(s-1,t)$ *for spacelike* $w$.

*Hence the reductions (i) and (ii) preserve* $s-t$ *and take* $(3,3)$ *to* $(2,2)$. *A reduction (iii) along a timelike direction lowers the mass, and one along a spacelike direction raises it; in particular* $(3,3)$ *reaches* $(3,1)$ *only by two timelike reductions, with* $M^2=m^2-\beta_1^2-\beta_2^2\le m^2$.

*Proof.* (i) is a standard result of two-time physics [8, 9]. (ii) Since $\{X\cdot\ell, P\cdot\ell\}=\ell\cdot\ell=0$, both constraints are first class. They restrict $X$ to the affine hyperplane $X\cdot\ell=c$ and $P$ to $\ell^\perp$, and they generate the gauge transformations $P\mapsto P+\varepsilon\ell$ and $X\mapsto X+\varepsilon'\ell$. The reduced phase space is therefore the cotangent bundle of an affine copy of $\ell^\perp/\mathbb{R}\ell$. The form induced on $\ell^\perp/\mathbb{R}\ell$ is nondegenerate of signature $(s-1,t-1)$, because $\ell$ together with a null $\ell'$ with $\ell\cdot\ell'=1$ spans a hyperbolic plane whose orthogonal complement maps isomorphically onto $\ell^\perp/\mathbb{R}\ell$; the mass-shell function $P\cdot P$ is gauge invariant and descends unchanged. (iii) The pair $X\cdot w$, $P\cdot w-\beta$ is second class, since $\{X\cdot w, P\cdot w\}=\epsilon\ne 0$, and its elimination leaves $T^*(w^\perp)$. Writing $P=P_\perp+\epsilon\beta w$ with $P_\perp\cdot w=0$, the mass shell becomes $P_\perp\cdot P_\perp+m^2+\epsilon\beta^2=0$. The last statement follows because every reduction other than (iii) along a timelike direction lowers $s$, and none raises it. □

In two-time physics the mass is a modulus of the embedding along the extra spacelike direction: in the massive gauge of Ref. [10], Eqs. (47)–(49), $P^{1'}=m$ (see also Ref. [42], Table 1). It is accommodated, and the theory does not fix its value. The $\mathfrak{sp}(2,\mathbb{R})$ constraints have nontrivial solutions only with at least two timelike dimensions, and with more than two the gauge symmetry no longer removes the ghosts, so the system exists physically only in signature $(d,2)$ (Refs. [8, 9]); in particular the reduction (i) of $(3,3)$ to $(2,2)$ retains ghosts.

**Corollary 13** (signature $(3,3)$)**.** *Let* $F=U^\perp\cong\mathbb{R}^2$ *be the normal space of a subspace* $U$ *of signature* $(3,1)$ *in flat* $\mathbb{R}^{3,3}$, *with* $\alpha=0$ *and a locally bounded* $W\le m_\Phi^2$. *Then* $L$ *is essentially self-adjoint on* $C_c^\infty(F)$, $\operatorname{spec} L$ *is unbounded below, and* $\operatorname{spec} L\subset(-\infty,m_\Phi^2]$. *The gauge reductions of Proposition 12 reach signature* $(3,1)$ *only by two timelike reductions, which lower the mass.*

*Proof.* By Lemma 10, $F$ is negative definite, so $L=m_\Phi^2-(-\Delta+m_\Phi^2-W)$ with $m_\Phi^2-W\ge 0$ locally bounded; Kato's theorem [32] gives essential self-adjointness, and Theorem 4(b),(c) gives the spectral statements. The last statement is Proposition 12. □

## 5 Flat para-Kähler space of signature $(4,4)$: Lagrangian sections and bilocal coordinates

Take $n=4$ in (1). By Sec. 2, a four-dimensional transverse section is the graph of an invertible linear map $\varphi: V_+\to V_-$, written below in the bases $\{\partial_{u_k}\}$ and $\{\partial_{v_k}\}$ as a $4\times4$ matrix.

**Proposition 14** (Lorentz group as stabilizer of a Lagrangian section)**.** *Let* $\Sigma=\{u+\varphi u: u\in V_+\}$ *with* $\varphi$ *invertible, and let* $S=1/2\,(\varphi+\varphi^T)$ *and* $A=1/2\,(\varphi-\varphi^T)$. *Then:*

(i) $h|_\Sigma$ *is represented by* $S$, *and* $\Sigma$ *is Lagrangian for* $\Omega=h(K\,\cdot,\cdot)$ *if and only if* $A=0$*;*

(ii) *for every $\Sigma$, $K: \Sigma \to K(\Sigma)$ reverses the induced metric, and $\Sigma$ is Lagrangian if and only if $\Sigma^{\perp} = K(\Sigma)$;*

(iii) *the stabilizer of $\Sigma$ in $Aut(h, K)$ is $\{\Lambda \in GL(4, \mathbb{R}): \Lambda^T \varphi \Lambda = \varphi\}$, the intersection of $O(S)$ with the stabilizer of $A$; if $S$ is nondegenerate, it equals $O(S) \cong O(h|_\Sigma)$ if and only if $\Sigma$ is Lagrangian; for $\varphi = \eta$ it is $O(3,1)$, acting on $\Sigma$ and on $N \equiv K(\Sigma)$ by the vector representation;*

(iv) *for $\varphi = \eta$, the space of Lorentz-equivariant linear maps $\Sigma \to N$ is $\mathbb{R} \cdot K|_\Sigma$, the commutant of $O(3,1)$ in $End(\mathbb{R}^{4,4})$ is isomorphic to $M_2(\mathbb{R})$, and the only invariant subspaces have dimension* 0, 4 *or* 8.

*Proof.* Use $h(\partial_{u_k}, \partial_{u_l}) = h(\partial_{v_k}, \partial_{v_l}) = 0$ and $h(\partial_{u_k}, \partial_{v_l}) = 1/2\, \delta_{kl}$. (i) $h(u + \varphi u, u' + \varphi u') = u^T S u'$. Since $K(u + \varphi u) = u - \varphi u$, $\Omega(u + \varphi u, u' + \varphi u') = u^T A u'$ up to sign. (ii) The first statement holds because $h(K\,\cdot, K\,\cdot) = -h$ on all of $\mathbb{R}^{4,4}$. Next, $K(\Sigma) = \mathrm{graph}(-\varphi)$, and $h(u + \varphi u, w - \varphi w) = -u^T A w$, which vanishes for all $u, w$ if and only if $A = 0$; since $\dim K(\Sigma) = \dim \Sigma^{\perp} = 4$, $K(\Sigma) \subseteq \Sigma^{\perp}$ is equivalent to equality. (iii) By (2), $\Lambda$ maps $u + \varphi u$ to $\Lambda u + \Lambda^{-T} \varphi u$, which lies in $\Sigma$ for all $u$ if and only if $\Lambda^{-T}\varphi = \varphi \Lambda$. Taking symmetric and antisymmetric parts, $\Lambda^T \varphi \Lambda = \varphi$ if and only if $\Lambda^T S \Lambda = S$ and $\Lambda^T A \Lambda = A$. If $S$ is nondegenerate, $A$ corresponds through $S$ to an element of $\mathfrak{so}(S)$, and invariance of $A$ under $O(S)$ makes it invariant under the adjoint action of the identity component, hence central in $\mathfrak{so}(S)$; since $\mathfrak{so}(S)$ is semisimple in dimension four, $A = 0$. For $\varphi = \eta$ the condition is the defining relation of $O(3,1)$, and $\Lambda$ acts on the parameter $u$, hence on $\Sigma$ and on $N$, by the vector representation. (iv) As an $O(3,1)$-module, $\mathbb{R}^{4,4} \cong \mathbb{R}^4 \otimes \mathbb{R}^2$ with $\mathbb{R}^4$ the vector representation, which is absolutely irreducible. Schur's lemma gives the commutant $\mathrm{End}(\mathbb{R}^2) = M_2(\mathbb{R})$ and $\mathrm{Hom}_{O(3,1)}(\Sigma, N) \cong \mathbb{R}$, spanned by $K|_\Sigma$; the invariant subspaces are $\mathbb{R}^4 \otimes U$ with $U \subseteq \mathbb{R}^2$. □

*Remark 15.* Proposition 14 is the flat, Lorentzian instance of a standard structure of Born geometry [19–21]. For $\varphi = \eta$, the metric $\mathcal{G} = h|_\Sigma \oplus (-h|_N)$, equal to $\eta \oplus \eta$ in the coordinates (9) below, is compatible with $(h, \Omega)$: $h^{-1}\mathcal{G}$ is a para-complex structure with eigenspaces $\Sigma$ and $N$, and $\Omega^{-1}\mathcal{G}$ is a complex structure. In the language of generalized metrics, $\Sigma = C_+$, $N = C_-$, $\mathcal{G}$ has signature (6,2), and the stabilizer of $\Sigma$ in $\mathrm{Aut}(h, K)$ is the diagonal subgroup of the double Lorentz group $O(3,1) \times O(3,1)$ of transformations preserving $h$ and $\mathcal{G}$ (Ref. [21], Sec. 2.4.1). In Ref. [20] the structure group of Born geometry, $O(d)$, refers to a Riemannian generalized metric (Eq. (3.24) there); the Lorentzian case is mentioned in the introduction of Ref. [19], and a generalized metric of signature $(2, 2(d-1))$ appears in Ref. [21]. The graph description of $\Sigma$ is the graph form of generalized metrics [43], Proposition 3.31; para-Hermitian and Born structures are treated systematically in Refs. [25, 44, 45]. Two further remarks justify the definite article in the title of Proposition 14. First, by part (iv) no $O(3,1)$-invariant subspace of signature (4,2) exists, so the Lorentz group of $\Sigma$ does not act on a (4,2) subspace as in two-time physics; shadows of the type of Ref. [8] arise from gauge reductions, which preserve $s - t$ (Proposition 12). Second, $\mathrm{GL}(4, \mathbb{R})$ also contains $SL(2, \mathbb{C})$, acting on $\mathbb{R}^4 \cong \mathbb{C}^2$ through the realified spinor representation; its invariant bilinear forms on $\mathbb{R}^4$ are antisymmetric, so by part (iii) it stabilizes only sections $\mathrm{graph}(\varphi)$ with $\varphi$ antisymmetric, on which $h$ vanishes by part (i). The Lorentz symmetry of a Lorentzian section is therefore realized by the vector representation. In signature (3,3), by contrast, $\mathrm{Aut}(h, K)$ contains no Lorentz group (Proposition 11), and the identity component of the stabilizer of the linear four-dimensional section of the companion paper is $SO(2) \times SO(1,1)$ (Theorem 15 there).

**Proposition 16** (bilocal dictionary)**.** *Let $\varphi = \eta$ and define coordinates $\sigma = (\sigma^1, \sigma^2, \sigma^3, \sigma^0)$ and $\nu = (\nu^1, \nu^2, \nu^3, \nu^0)$ on $\mathbb{R}^{4,4}$ by*

$$\sigma = (x_1, x_2, x_3, \tau_4), \qquad \nu = (\tau_1, \tau_2, \tau_3, x_4). \tag{9}$$

*Then*

$$h = \eta(d\sigma, d\sigma) - \eta(dv, dv), \qquad K(\sigma, v) = (v, \sigma), \tag{10}$$

*$\Sigma = \{v = 0\}$ and $N = \{\sigma = 0\}$. With $z_1^\mu = \sigma^\mu + v^\mu$ and $z_2^\mu = \sigma^\mu - v^\mu$, $h = \eta_{\mu\nu}\, dz_1^\mu\, dz_2^\nu$, the product of differentials being symmetrized, so that $h\left(\partial_{z_1^\mu}, \partial_{z_2^\nu}\right) = 1/2\, \eta_{\mu\nu}$; each leaf $\{z_2 = const\}$ and $\{z_1 = const\}$ is totally null, $K = +1$ on the first family and $-1$ on the second, and $O(3,1) \subset Aut(h, K)$ acts diagonally, $z_{1,2} \mapsto \Lambda z_{1,2}$. Moreover $\Box_\sigma - \Box_v = 4\eta^{\mu\nu}\partial_{z_1^\mu}\partial_{z_2^\nu}$. The normal directions $v^1, v^2, v^3$ are h-timelike, and $v^0 = x_4$, with $\partial_{x_4} = K\partial_{\tau_4}$, is h-spacelike.*

*Proof.* Substituting (9) into (1) gives (10), since $\eta$ has its negative entry in the fourth slot; $K\,\partial_{x_k} = \partial_{\tau_k}$ gives $K(\sigma, v) = (v, \sigma)$. A vector $u + \eta u$ of $\Sigma$ equals $\sum_{k\le 3} u_k\, \partial_{x_k} + u_4\, \partial_{\tau_4}$, so $\Sigma = \{v = 0\}$. In the coordinates of Sec. 2, $z_1^\mu = u_\mu$ and $z_2^\mu = \eta^{\mu\mu} v_\mu$ (no sum), so $\eta_{\mu\nu} dz_1^\mu dz_2^\nu = \sum_k d\,u_k\, dv_k = h$. By (2), $\Lambda \in O(3,1)$ acts by $u \mapsto \Lambda u$ and $v \mapsto \Lambda^{-T} v = \eta\Lambda\eta\, v$, that is, $z_2 \mapsto \Lambda z_2$. The last identity follows from $\partial_\sigma = \partial_{z_1} + \partial_{z_2}$ and $\partial_v = \partial_{z_1} - \partial_{z_2}$. □

In double-field-theory notation [23, 24], $z_1 = x$ and $z_2 = \eta^{-1}\tilde{x}$, so that $h = dx^\mu d\tilde{x}_\mu$ is the $O(4,4)$ metric, $\sigma$ and $v$ are the combinations $1/2\,(x \pm \eta^{-1}\tilde{x})$, and $h^{AB}\,\partial_A\,\partial_B = \Box_\sigma - \Box_v$ is proportional to $\partial_x \cdot \partial_{\tilde{x}}$, the operator of the level-matching constraint. We read $z_1$ and $z_2$ as two events because the Poincaré group of $\Sigma$ acts on them diagonally; the metric $h$, which vanishes on each copy, does not single out this reading, and the dynamics below is not that of a two-particle system. With this caveat, in the minimal para-Kähler space in which $\mathrm{Aut}(h, K)$ realizes the Lorentz group as the stabilizer of a section, the three timelike normal directions are internal relative coordinates of a two-point object, and their timelike character is the sign of the relative metric imposed by $h(K\,\cdot, K\,\cdot) = -h$. The observed time is the bulk-timelike direction $\tau_4 \in \Sigma$, and its partner $\partial_{x_4} = K\,\partial_{\tau_4}$ is the relative time. This places the construction in the territory of bilocal fields [26, 46], Born reciprocity [47] and the covariant oscillator [28–30]. A curved analogue of the metric $\eta(dz_1, dz_2)$ on pairs of points is the kinematic space of conformal field theory [48, 49].

**Corollary 17** (free mass formula)**.** *A free bulk field $(h^{AB}\partial_A\partial_B - m_\Phi^2)\Phi = 0$ with momentum $p$ along $\Sigma$ and $\pi$ along $N$ has*

$$M^2 = m_\Phi^2 - \eta(\pi, \pi) = m_\Phi^2 + \pi_{x_4}^2 - \sum_{i=1}^{3} \pi_{\tau_i}^2. \tag{11}$$

*Proof.* By (10), $h^{-1} = \eta^{-1} \oplus (-\eta^{-1})$, so $\Box_\sigma\Phi = (m_\Phi^2 + \Box_v)\Phi$, and $\Box_v e^{i\pi\cdot v} = -\eta(\pi, \pi)e^{i\pi\cdot v}$. □

Equation (11) is Theorem 4(a) for the fiber $N$ of signature (1,3): the gradient along $x_4$ is a spacelike gradient contribution and those along $\tau_1, \tau_2, \tau_3$ are temporal contributions in the sense of Definition 2. The three timelike normal directions lower $M^2$, and only the partner of the observed time raises it. Restricting the normal momenta to the closed light cone $\{\pi: \eta(\pi, \pi) \le 0\}$, a Lorentz-invariant condition, gives $M^2 = m_\Phi^2 + \mu^2 \ge m_\Phi^2$ with $\mu^2 = -\eta(\pi, \pi)$; the spectrum is then continuous with infinite multiplicity, and the projector onto the cone is nonlocal along $N$.

**The reading with the generalized metric.** The same space carries a second Lorentz-invariant metric, the generalized metric $\mathcal{G} = \eta \oplus \eta$ of Remark 15. If $\mathcal{G}$ rather than $h$ is taken as the kinetic metric, the free mass formula becomes $M^2 = m_\Phi^2 + \eta(\pi, \pi) = m_\Phi^2 + \sum_i \pi_{\tau_i}^2 - \pi_{x_4}^2$: the directions $\tau_i$ are then spacelike and raise $M^2$, and the relative time lowers it. Which normal directions count as temporal is fixed by the choice of kinetic metric, and in either reading the sign of a contribution to $M^2$ is the sign of the kinetic metric on the corresponding direction. In this paper $h$ is the kinetic metric, because the question concerns directions that are timelike for the para-Hermitian metric.

# 6 A constrained bilocal oscillator: model $A_K$

**Definition 18** (model $A_K$). On $(\mathbb{R}^{4,4}, h, K)$ in the coordinates (9), let

$$S = \int d^4\sigma\, d^4v \left[-1/2\, h^{AB}\, \partial_A\Phi\, \partial_B\Phi - 1/2\left(m_\Phi^2 + \omega^2 h(v,v)\right)\Phi^2\right], \qquad \omega > 0, \tag{12}$$

and define the physical subspace by the subsidiary condition

$$C^{ab}\Phi \equiv \left(p^a A^b - p^b A^a\right)\Phi = 0, \qquad p_a = -i\,\partial_{\sigma^a}, \quad A^b = \omega v^b + \eta^{bc}\,\partial_{v^c}, \tag{13}$$

where indices on $N$ are carried to $\Sigma$ by $K$, the Lorentz-equivariant map of Proposition 14 (iv). The condition is local, since $p$ acts as a derivative along $\Sigma$.

The potential $\omega^2 h(v,v) = -\omega^2\eta(v,v)$ is Lorentz invariant and breaks only the translations along $N$; it fixes the origin $v = 0$, the coincidence of the two points, and it is unbounded below, so that the bulk mass function $W = m_\Phi^2 + \omega^2 h(v,v)$ is tachyonic at large relative spatial separation. The structure $K$ fixes $N = K(\Sigma)$, the identification of indices that makes (13) local, and the sign pattern $h|_N = -\eta$. The field equation is $\Box_\sigma\Phi = (m_\Phi^2 + H_N)\Phi$ with

$$H_N = \Box_v - \omega^2\eta(v,v) = O_{v^0} - \sum_{i=1}^{3} O_{v^i}, \qquad O_y = -\partial_y^2 + \omega^2 y^2. \tag{14}$$

For $p \in \mathbb{R}^{3,1}$ let

$$\mathcal{K}_p = \{\chi \in L^2(\mathbb{R}^4, d^4v)\colon\ (m_\Phi^2 + H_N)\chi = -\eta(p,p)\,\chi,\ \ C^{ab}(p)\chi = 0 \text{ for all } a,b\}, \tag{15}$$

with the equations understood in the sense of distributions; $e^{ip\cdot\sigma}\chi$ is then a physical plane-wave solution. Let $\psi_n$ denote the normalized Hermite functions of frequency $\omega$.

**Theorem 19** (free spectrum).

*(i)* *For spacelike $p$ and for null $p \neq 0$, $\mathcal{K}_p = 0$.*

*(ii)* *For timelike $p$, $\mathcal{K}_p \neq 0$ if and only if $-\eta(p,p) = M_n^2$ for some $n \in \{0,1,2,\dots\}$, where*

$$M_n^2 = m_\Phi^2 + 2\omega(n-1); \tag{16}$$

*then $\mathcal{K}_p = \mathbb{C}\,\chi_n(\cdot;\hat p)$, where $\hat p = p/M_n$ and $\chi_n(v;\hat p) = \chi_n(\Lambda^{-1}v)$ for any orthochronous Lorentz transformation $\Lambda$ with $\Lambda(\mathbf{0},1) = \pm\hat p$, $\chi_n = \psi_n(v^0)\prod_{i=1}^3 \psi_0(v^i)$ being the rest-frame profile; since $\chi_n$ is invariant under rotations, $\chi_n(\cdot;\hat p)$ does not depend on the choice of $\Lambda$.*

*(iii)* *$\mathcal{K}_0 = 0$ if and only if $m_\Phi^2/(2\omega) \notin \mathbb{Z}$; otherwise $\mathcal{K}_0$ is infinite-dimensional.*

*Consequently, if $m_\Phi^2/(2\omega) \notin \mathbb{Z}$, the positive-energy physical one-particle space is the direct sum, over the $n$ with $M_n^2 > 0$, of the scalar unitary irreducible representations of the Poincaré group of $\Sigma$ with mass $M_n$, each occurring once and carrying positive Klein–Gordon norm. Levels with $M_n^2 \leq 0$ do not occur.*

*Proof. Algebra.* With $A^{b\dagger} = \omega v^b - \eta^{bc}\,\partial_{v^c}$ one finds $[A^a, A^b] = 0$, $[A^a, A^{b\dagger}] = 2\omega\eta^{ab}$, $H_N = -\eta_{ab}A^{a\dagger}A^b - 4\omega$ and $[H_N, A^b] = 2\omega A^b$. Thus $A^i = \omega v^i + \partial_{v^i}$ annihilates the $v^i$ oscillator, and $A^0 = \omega v^0 - \partial_{v^0}$ is the creation operator of the $v^0$ oscillator. Since $[H_N, C^{ab}] = 2\omega C^{ab}$, the kernel of $C(p)$ is invariant under $H_N$. Both $H_N$ and the family $C(p)$ are Lorentz covariant, $C^{ab}(\Lambda p) = \Lambda^a_c\Lambda^b_d\, U(\Lambda)C^{cd}(p)U(\Lambda)^{-1}$ with $(U(\Lambda)\chi)(v) = \chi(\Lambda^{-1}v)$, and $U$ is unitary on $L^2(\mathbb{R}^4, d^4v)$ because $d^4v$ is invariant; hence $\mathcal{K}_{\Lambda p} = U(\Lambda)\mathcal{K}_p$, and it suffices to treat one representative of each orbit of $p$. Since $C(p)$ is linear in $p$, also $\mathcal{K}_{-p} = \mathcal{K}_p$.

*Spacelike p.* Take $p = (p^1, 0,0,0)$ with $p^1 \neq 0$. Then $C^{10} = p^1 A^0$, so $(\omega v^0 - \partial_{v^0})\chi = 0$, whose distributional solutions are $\chi = e^{\omega (v^0)^2/2} g(v^1, v^2, v^3)$; such $\chi$ is square integrable only if $g = 0$.

*Null* $p \neq 0$. Take $p^1 = p^0 \neq 0$, $p^2 = p^3 = 0$. Then $C^{10} \propto A^0 - A^1$ and $C^{20}, C^{30} \propto A^2, A^3$. In the variables $U = v^0 - v^1$, $V = v^0 + v^1$ the first condition reads $(\omega U - 2\,\partial_V)\chi = 0$, so $\chi = e^{\omega UV/2} f(U, v^2, v^3)$ and $|\chi|^2 = e^{\omega UV}|f|^2$. For almost every $U \neq 0$ the $V$-integral diverges unless $f(U,\cdot) = 0$, so $\chi = 0$.

*Timelike p.* In the rest frame $p = (\mathbf{0}, p^0)$, $C^{ij} = 0$ and $C^{i0} = -p^0 A^i$, so $\left(\omega v^i + \partial_{v^i}\right)\chi = 0$ for $i = 1,2,3$, whose square-integrable solutions are $\chi = g(v^0) \prod_i \psi_0\left(v^i\right)$ with $g \in L^2(\mathbb{R})$. On such $\chi$, $H_N \chi = [(\mathcal{O}_{v^0} g) - 3\omega g] \prod_i \psi_0$, and the square-integrable distributional solutions of $\mathcal{O}_{v^0} g = \lambda g$ are $g \propto \psi_n$ with $\lambda = \omega(2n+1)$: pairing the equation with the Hermite functions, which are Schwartz functions, eigenfunctions of $\mathcal{O}_{v^0}$ and an orthonormal basis of $L^2(\mathbb{R})$, shows that $g$ has no component along $\psi_m$ unless $\lambda = \omega(2m+1)$. Hence $\mathcal{K}_p \neq 0$ exactly when $-\eta(p,p) = m_\Phi^2 + \omega(2n+1) - 3\omega = M_n^2$, and then it is spanned by $\chi_n$.

$p = 0$. The condition (13) is empty. The products of Hermite functions form an orthonormal basis of $L^2(\mathbb{R}^4)$ of eigenfunctions of $H_N$ with eigenvalues $\omega(2n_0+1) - \omega \sum_i (2n_i + 1) \in 2\omega\mathbb{Z}$, each value occurring with infinite multiplicity; pairing as above, a nonzero square-integrable solution of $(m_\Phi^2 + H_N)\chi = 0$ exists if and only if $-m_\Phi^2 \in 2\omega\mathbb{Z}$, and then the solutions form an infinite-dimensional space.

*Representations.* The rest-frame profile $\chi_n$ is invariant under rotations, so each level carries spin zero, and the covariance $\mathcal{K}_{\Lambda p} = U(\Lambda)\mathcal{K}_p$ makes the space of positive-energy wave packets $\Phi = \int d^3k\,(2E_n)^{-1} a(\mathbf{k})\, e^{ip\cdot\sigma} \chi_n(v;\hat{p})$, with $p = (\mathbf{k}, E_n)$ on the mass shell $M_n$ and $a \in \mathcal{S}(\mathbb{R}^3)$, unitarily equivalent to the scalar representation of mass $M_n$. Since $\|\chi_n(\cdot;\hat{p})\| = \|\chi_n\|$, the Klein–Gordon norm of such a packet is $(2\pi)^3 \int d^3k\,(2E_n)^{-1} |a(\mathbf{k})|^2 \|\chi_n\|^2 > 0$. Negative-energy solutions are the complex conjugates of these, because $\chi_n$ is real. Levels with $M_n^2 \leq 0$ would require spacelike, null or zero $p$, which part (i) and the hypothesis exclude. □

*Remark 20.* The exclusion of levels with $M_n^2 \leq 0$ is a consequence of the subsidiary condition. If $m_\Phi^2 > 2\omega$ all levels are present and the gap is $M_0^2 = m_\Phi^2 - 2\omega$. If $m_\Phi^2/(2\omega) \in \mathbb{Z}$, for instance $m_\Phi = 0$, the hypothesis of the last statement of Theorem 19 fails: the levels with $p \neq 0$ are still the $M_n^2 > 0$ (for $m_\Phi = 0$, $M^2 = 2\omega, 4\omega, \ldots$), but $\mathcal{K}_0$ is infinite-dimensional and adds solutions independent of $\sigma$. In a boosted frame the profile reads $\chi_n(v;\hat{p}) = c_n H_n\left(\sqrt{\omega}\,\zeta\right) e^{-\omega Q/2}$ with $\zeta = -\eta(v,\hat{p})$ and the positive-definite form $Q = \eta(v,v) + 2\eta(v,\hat{p})^2$, where $c_n$ is the normalization of $\chi_n$ (Online Resource 1, Sec. S2.4).

**Comparison with the covariant oscillator.** Table 1 compares $A_K$ with the covariant harmonic oscillator of Kim, Noz and Oh [30], which follows Feynman, Kislinger and Ravndal [28] and whose relative metric is $+\eta$. In terms of $P = -i\,\partial_v$ and $X = v$, $H_N = -(P\cdot P + \omega^2 X \cdot X)$ with the dot product $\eta$, so that in $A_K$ quanta of the relative time raise $M^2$ and quanta of the relative spatial separation lower it; in the covariant oscillator the roles are reversed, and the subsidiary condition of Ref. [30], Eq. (9), freezes the relative time where (13) freezes the relative spatial separation. Up to normalization, $H_N$ is the negative of the covariant-oscillator operator of Refs. [28, 30]. The internal space $L^2(\mathbb{R}^4, d^4v)$ with the unitary action $U(\Lambda)$ is the representation space of Dirac's Lorentz-covariant four-dimensional oscillator [50], in which the temporal and the spatial oscillators enter the invariant energy with opposite signs, as they enter $H_N$. The para-Kähler structure enters $A_K$ only through the sign of $h$ on $N$ and through the identification $K$ of $N$ with $\Sigma$, which by Proposition 14 (iv) is the unique Lorentz-equivariant one up to scale: $A_K$ is a covariant oscillator on the flat bulk $\eta \oplus (-\eta)$ with the opposite internal sign and the complementary subsidiary condition. Bars [51], Sec. IV, Eqs. (4.10) and (4.21), classifies the Fock vacua of the covariant oscillator and finds a ghost-free sector, whose vacuum is not Lorentz invariant, in which each timelike quantum lowers the invariant. The tower of $A_K$ consists of excitations of the relative time with the relative spatial separation frozen. In the covariant-oscillator literature the excitations of the relative time are removed by the subsidiary condition [30]; here they carry the whole tower, as a consequence of the sign $h|_N = -\eta$.

Table 1: Model $A_K$ and the covariant oscillator of Kim, Noz and Oh [30]. Equation numbers refer to Ref. [30]; $\Omega_0$ is its oscillator constant, set to 1 there, and the multiplicity is computed from its Eq. (18).

| | Kim–Noz–Oh | $A_K$ |
|---|---|---|
| Relative metric | $+\eta$ | $-\eta$, from $h(K\,\cdot, K\,\cdot) = -h$ |
| Frozen by the condition | relative time, Eq. (9) | relative spatial separation ($\tau_i$) |
| Excited | relative spatial separation | relative time ($x_4$) |
| Levels | $M^2 = m_0^2 + \Omega_0(j+1)$, Eq. (21) | $M_n^2 = m_\Phi^2 + 2\omega(n-1)$ |
| Multiplicity | $(j+1)(j+2)/2$ | 1, spin 0 |

## 6.1 Conserved quantities and the Cauchy problem

**Lemma 21** (orthogonality of physical profiles). *For timelike unit vectors $\hat{p}, \hat{p}'$ and $n \neq m$,* $\langle \chi_n(\cdot;\hat{p}), \chi_m(\cdot;\hat{p}')\rangle_{L^2(\mathbb{R}^4)} = 0$.

*Proof.* Since $\chi_n(\cdot;\hat{p})$ is a Lorentz image of $\chi_n$ and $H_N$ is Lorentz invariant, $H_N\chi_n(\cdot;\hat{p}) = 2\omega(n-1)\chi_n(\cdot;\hat{p})$, and $\chi_n(\cdot;\hat{p}) \in \mathcal{S}(\mathbb{R}^4)$. The operator $H_N$ is symmetric on $\mathcal{S}(\mathbb{R}^4)$, and eigenvectors of a symmetric operator for different eigenvalues are orthogonal. □

Let the physical solutions be the real fields

$$\begin{aligned} \Phi(\sigma,v) &= \sum_{n\in\mathcal{N}} \int \frac{d^3k}{(2\pi)^3\, 2E_n(\mathbf{k})} \left[a_n(\mathbf{k})\, e^{ip_n\cdot\sigma}\chi_n(v;\hat{p}_n) + \mathrm{c.c.}\right], \\ p_n &= (\mathbf{k}, E_n), \qquad E_n = \sqrt{|\mathbf{k}|^2 + M_n^2}, \end{aligned} \tag{17}$$

with $\mathcal{N} = \{n : M_n^2 > 0\}$, the $\chi_n$ normalized, and $a_n \in \mathcal{S}(\mathbb{R}^3)$, finitely many of them nonzero. We write $\phi_n(\sigma) = \int d^3k\, ((2\pi)^3\, 2E_n)^{-1}\left[a_n e^{ip_n\cdot\sigma} + \mathrm{c.c.}\right]$ for the four-dimensional Klein–Gordon field of mass $M_n$ with the coefficients $a_n$, and $\mathcal{H}_{\mathrm{phys}}$ for the completion of the space of solutions (17) in the norm $E^{1/2}$ of Proposition 22.

**Proposition 22** (energy and evolution on the physical space). *Assume $m_\Phi^2/(2\omega) \notin \mathbb{Z}$.*

*(i) The canonical energy of (12),*

$$\begin{aligned} E[\Phi] = \frac{1}{2}\int_{\mathbb{R}^3\times\mathbb{R}^4} &[(\partial_{\sigma^0}\Phi)^2 + |\nabla_{\boldsymbol{\sigma}}\Phi|^2 + (\partial_{v^0}\Phi)^2 - \sum_{i=1}^{3}\left(\partial_{v^i}\Phi\right)^2 \\ &+ (m_\Phi^2 + \omega^2(v^0)^2 - \omega^2|\mathbf{v}|^2)\Phi^2]\, d^3\boldsymbol{\sigma}\, d^4v, \end{aligned} \tag{18}$$

*a local functional of the Cauchy data $(\Phi, \partial_{\sigma^0}\Phi)$ on $\{\sigma^0 = t\} \times N$, is conserved along solutions that decay sufficiently fast in $\boldsymbol{\sigma}$ and $v$. As a functional of the Cauchy data it is unbounded above and below.*

*(ii) On the physical solutions (17),*

$$E[\Phi] = \sum_{n\in\mathcal{N}} \int \frac{d^3k}{(2\pi)^3\, 2E_n(\mathbf{k})}\, E_n(\mathbf{k})\, |a_n(\mathbf{k})|^2, \tag{19}$$

*which is positive for $\Phi \neq 0$. The conserved Klein–Gordon product $\langle \Phi_1, \Phi_2 \rangle_{KG} = i \int \big(\overline{\Phi_1}\, \partial_{\sigma^0} \Phi_2 - \partial_{\sigma^0} \overline{\Phi_1}\, \Phi_2\big)\, d^3\boldsymbol{\sigma}\, d^4 v$ of the positive-frequency part $\Phi^{(+)}$ of (17), the part containing the $a_n$, is $\langle \Phi^{(+)}, \Phi^{(+)} \rangle_{KG} = \sum_{n\in\mathcal{N}} \int d^3k\, ((2\pi)^3\, 2E_n)^{-1} |a_n(\boldsymbol{k})|^2$, positive and diagonal in $n$; for the real field itself it vanishes.*

*(iii) Let $\mathcal{A} = \oplus_{n\in\mathcal{N}} L^2(\mathbb{R}^3, d^3k/2(2\pi)^3)$. The map $\Phi \mapsto (a_n)_{n\in\mathcal{N}}$ extends to a unitary map from $\mathcal{H}_{phys}$ onto $\mathcal{A}$; translation in $\sigma^0$ extends to a strongly continuous unitary group on $\mathcal{H}_{phys}$; and $\mathcal{H}_{phys}$ embeds continuously and injectively into the space of Cauchy data, with*

$$\|\Phi(t)\|_{L^2(\mathbb{R}^3\times\mathbb{R}^4)} \leq \sqrt{2}\, E[\Phi]^{1/2}/M_{\min}, \qquad \|\,\partial_{\sigma^0}\Phi(t)\|_{L^2(\mathbb{R}^3\times\mathbb{R}^4)} \leq \sqrt{2}\, E[\Phi]^{1/2}, \tag{20}$$

*where $M_{min} = min_{n\in\mathcal{N}} M_n$.*

*(iv) A solution $\Phi \in C^2\big(\mathbb{R}; L^2(\mathbb{R}^3 \times \mathbb{R}^4)\big)$ of the field equation, in the sense of distributions, is determined by $(\Phi, \partial_{\sigma^0}\Phi)$ at one time. If $C^{ab}\Phi$ and $\partial_{\sigma^0} C^{ab}\Phi$ vanish at one time, $C^{ab}\Phi$ vanishes at all times.*

*Proof.* (i) The energy density is the canonical $T^0_0$ of (12), with $h^{-1} = \eta^{-1} \oplus (-\eta^{-1})$ and $\omega^2 h(v,v) = \omega^2 (v^0)^2 - \omega^2 |\mathbf{v}|^2$; its conservation along solutions that decay sufficiently fast, such as the solutions (17), follows from the field equation by integration by parts. Unboundedness: take $\partial_{\sigma^0}\Phi = 0$ and $\Phi = \phi(\boldsymbol{\sigma})\,\chi$ with $\chi = \psi_0(v^0)\,\psi_k(v^1)\,\psi_0(v^2)\,\psi_0(v^3)$; then the integral over $\boldsymbol{\sigma}$ and $v$ of the $v$-gradient and potential terms of (18) equals $\langle \chi, (m_\Phi^2 + H_N)\chi \rangle \int \phi^2\, d^3\boldsymbol{\sigma} = \big(m_\Phi^2 - 2\omega(k+1)\big) \int \phi^2\, d^3\boldsymbol{\sigma}$, which tends to $-\infty$ as $k \to \infty$; profiles excited along $v^0$ give $+\infty$.

(ii) On (17), integration over $\boldsymbol{\sigma}$ pairs $\mathbf{k}$ with $\pm\mathbf{k}'$, and integration by parts in $v$ turns the $v$-gradient and potential terms into $\langle \chi, (m_\Phi^2 + H_N)\chi' \rangle$. Each $\chi_n(\cdot; \hat{p})$ is an eigenfunction of $m_\Phi^2 + H_N$ with eigenvalue $M_n^2$, and by Lemma 21 all terms pairing different $n$ vanish. For equal $n$, the terms pairing $\mathbf{k}$ with $-\mathbf{k}$ oscillate as $e^{\mp 2iE_n t}$ and carry the coefficient $(-E_n^2 + |\mathbf{k}|^2 + M_n^2)\langle \chi_n(\cdot; \hat{p}_n(\mathbf{k})), \chi_n(\cdot; \hat{p}_n(-\mathbf{k}))\rangle = 0$, while the terms pairing $\mathbf{k}$ with itself carry $E_n^2 + |\mathbf{k}|^2 + M_n^2 = 2E_n^2$; this gives (19). The Klein–Gordon product of the positive-frequency parts is treated in the same way: only equal $\mathbf{k}$ can be used, and the terms with different $n$ vanish by Lemma 21.

(iii) By (19), $E[\Phi] = 1/2\,(2\pi)^{-3} \sum_n \|\, a_n\|_{L^2}^2$, so the map is isometric on (17), and its range is dense; translation by $t$ multiplies $a_n$ by $e^{-iE_n t}$. In the Fourier transform in $\boldsymbol{\sigma}$, $\widehat{\Phi}(\mathbf{k},\cdot,t) = \sum_n (2E_n)^{-1} \big[a_n(\mathbf{k}) e^{-iE_n t} \chi_n(\cdot; \hat{p}_n(\mathbf{k})) + \overline{a_n(-\mathbf{k})} e^{iE_n t} \chi_n(\cdot; \hat{p}_n(-\mathbf{k}))\big]$. By Lemma 21 the terms with different $n$ are orthogonal in $L^2(\mathbb{R}^4)$, so $\|\widehat{\Phi}(\mathbf{k},\cdot,t)\|^2 \leq \sum_n (|a_n(\mathbf{k})|^2 + |a_n(-\mathbf{k})|^2)/(2E_n^2)$, and integration over $\mathbf{k}$ gives the first bound in (20); the second follows in the same way. Injectivity: if $P_n$ is the orthogonal projection onto $\chi_n(\cdot; \hat{p}_n(\mathbf{k}))$, Lemma 21 and the computation above give

$$P_n\big(E_n \widehat{\Phi} + i\,\partial_{\sigma^0}\widehat{\Phi}\big)(\mathbf{k},\cdot,t) = a_n(\mathbf{k})\, e^{-iE_n t}\, \chi_n(\cdot; \hat{p}_n(\mathbf{k})),$$

so the data determine every $a_n$, and the bound extends to $\mathcal{H}_{\text{phys}}$ by continuity.

(iv) Take the Fourier transform in $\boldsymbol{\sigma}$ and expand in the products $\psi_\alpha = \psi_{n_0}(v^0) \prod_i \psi_{n_i}(v^i)$. The coefficients $c_\alpha(\mathbf{k}, t)$ satisfy, for almost every $\mathbf{k}$, $\partial_t^2 c_\alpha = -(|\mathbf{k}|^2 + m_\Phi^2 + \lambda_\alpha) c_\alpha$ with $\lambda_\alpha = 2\omega(n_0 - n_1 - n_2 - n_3 - 1)$, a linear ordinary differential equation of second order, so vanishing data give $c_\alpha \equiv 0$. The operators $C^{ab}$ commute with $\square_\sigma$, and $[H_N, C^{ab}] = 2\omega C^{ab}$; hence $\Psi = C^{ab}\Phi$ satisfies $\Box_\sigma \Psi = (m_\Phi^2 - 2\omega + H_N)\Psi$. The Hermite–Fourier coefficients of $\Psi$ are finite linear combinations of those of $\Phi$ and $\partial_{\sigma^0}\Phi$, and they satisfy the same kind of equation, so vanishing data give $\Psi \equiv 0$. □

Proposition 22 establishes well-posed unitary evolution on the constrained physical space in its energy norm. In the rest-frame description of the physical sector, the timelike normal directions have no independent oscillator excitations: every $\tau$-oscillator is in its ground state, and $\sigma^0$ is the only evolution parameter. The sector is selected by the local condition (13) together with square integrability along $N$, a global condition that plays the role of the nonlocal constraints of Ref. [17], and, if $m_\Phi^2 \le 2\omega$, by the exclusion of the levels with $M_n^2 \le 0$. Off the kernel of $C$ the energy conjugate to $\sigma^0$ is unbounded below, and the unconstrained Cauchy problem is ill posed: the coefficients $c_\alpha$ of part (iv) grow at the rates $(-\lambda_\alpha - |\mathbf{k}|^2 - m_\Phi^2)^{1/2}$, which are unbounded as the number $n_1 + n_2 + n_3$ of quanta along the timelike normal directions grows. Part (iii) identifies $\mathcal{H}_{\text{phys}}$ with a subspace continuously embedded in the $L^2$ Cauchy data on $\{\sigma^0 = t\} \times N$; we have not characterized this subspace by local conditions on the data. Finite propagation speed holds for each $\phi_n$ and for $\Phi(\sigma, 0) = \sum_n \chi_n(0)\phi_n(\sigma)$, since $\chi_n(0;\hat{p})$ does not depend on $\hat{p}$, but we do not claim it for $\Phi(\sigma, \nu)$ at $\nu \neq 0$. The results hold at the free level only (Sec. 7).

*Remark 23* (two first-class constraints). An alternative imposes two mass-shell constraints, $C_h = h^{-1}(P,P) + m_\Phi^2$ and $C_\mathcal{G} = \mathcal{G}^{-1}(P,P) + m_\mathcal{G}^2$, with $\mathcal{G} = \eta \oplus \eta$ the generalized metric of Remark 15, of signature (6,2), and $P = (p, \pi)$ the momenta along $\Sigma$ and $N$. Both depend on $P$ only, so they Poisson-commute, and their sum and difference give $M^2 = 1/2\left(m_\Phi^2 + m_\mathcal{G}^2\right)$ and $\mu^2 \equiv -\eta(\pi,\pi) = 1/2\left(m_\mathcal{G}^2 - m_\Phi^2\right)$. The single mass level then carries all normal momenta on the quadric $\eta(\pi,\pi) = -\mu^2$, which consists of two copies of hyperbolic space $H^3$ if $\mu^2 > 0$, a one-sheeted hyperboloid if $\mu^2 < 0$ and the null cone if $\mu^2 = 0$; there is no tower, and the level has infinite multiplicity, so (P2) fails. Adding potentials $f(\nu)$ to $C_h$ and $g(\nu)$ to $C_\mathcal{G}$ gives $\{C_h + f, C_\mathcal{G} + g\} = 2\eta^{ab}\pi_b\, \partial_a(f+g)$, which vanishes on the whole constraint surface if and only if $\nabla(f+g) = 0$; then $C_h + C_\mathcal{G}$ is again independent of $\nu$, and $M^2$ takes a single value. In the frame $z_1 = x$, $z_2 = \eta^{-1}\tilde{x}$ of Proposition 16, $C_\mathcal{G}$ and $C_h$ are, up to normalization, the Hamiltonian and level-matching constraints of the metaparticle [52], Eqs. (5) and (6); the observed momentum there is the momentum conjugate to $x$, whereas here it is the momentum conjugate to $\sigma$.

# 7 Interactions and metric fluctuations

## 7.1 Criteria

Model $A_K$ satisfies (P3) for the scalar sector, (P5) in the weak sense of Sec. 2.3 (the bulk violates chronology), and (P7). It satisfies (P1) on the levels present, (P4) only in the restricted sense of Proposition 22, and (P6) with explicit breaking: the origin $\nu = 0$ is preferred and the normal energies are not conserved. The Einstein–Hilbert extension of Proposition 25 fails (P3), since its metric sector contains ghosts. The scalar model fails the direct temporal-origin criterion (P2). In the splitting of $N$ adapted to the momentum, $T_+ = \mathbb{R}\,K\hat{p}$, the decomposition (6) of the level $M_n^2$ has

$$\begin{aligned} \mathcal{T} &= -\sum_{i=1}^{3} \|\,\partial_{\nu^i}\psi_0\|^2 = -3/2\,\omega, \\ \mathcal{S} &= \|\,\partial_{\nu^0}\psi_n\|^2 = 1/2\,\omega(2n+1), \\ \mathcal{W} &= m_\Phi^2 + 1/2\,\omega(2n+1) - 3/2\,\omega, \end{aligned} \tag{21}$$

and $\mathcal{B} = 0$, with sum $M_n^2$. The positive part of $M_n^2 - m_\Phi^2$ is carried by the spacelike gradient along $x_4$ and by the input potential; the temporal contribution is negative.

### 7.2 Interactions

**Proposition 24** (local vertices). *Let $\psi_k$ be the normalized Hermite functions of frequency $\omega$ in one $\tau$ direction. The first-order matrix element of the vertex $g\int \Phi^3$ between two physical states at rest and a state with $2k$ quanta in one $\tau$ direction is proportional to*

$$I_{2k} = \int_{\mathbb{R}} \psi_0\,(y)^2 \psi_{2k}(y)\,dy, \qquad I_0 = \frac{\sqrt{6}}{3\pi^{1/4}}\,\omega^{1/4}, \qquad \frac{I_{2k+2}}{I_{2k}} = -\frac{1}{3}\sqrt{\frac{2k+1}{2k+2}}, \tag{22}$$

*with a nonzero factor for suitable quantum numbers along $\nu^0$, and $I_{2k+1} = 0$. Such states solve the unconstrained field equation with $H_N$-eigenvalue $2\omega(n_0 - n_\tau - 1)$, where $n_\tau$ is the total number of $\tau$ quanta, and they violate (13). Hence $\int \Phi^3$ does not leave the physical subspace invariant, and neither does $\int \Phi^k$ for any $k \geq 3$. The sourced states have $M^2 = m_\Phi^2 + 2\omega(n_0 - n_\tau - 1)$, which is negative for large $n_\tau$, so enlarging the state space to accommodate them violates (P1).*

*Proof.* For two physical states at rest with profiles $\chi_{n_1}$ and $\chi_{n_2}$ and a third state with profile $\psi_m(\nu^0)\,\psi_{2k}(\nu^1)\,\psi_0(\nu^2)\,\psi_0(\nu^3)$, the $\nu$-integral of the product of the three profiles factorizes as

$$\int_{\mathbb{R}} \psi_{n_1}\,\psi_{n_2}\psi_m\,d\nu^0 \cdot I_{2k}\,I_0^2,$$

and the first factor is nonzero, for instance, for $m = n_1 = n_2 = 0$. The overlaps are computed in Appendix A, which also gives the corresponding overlap for $\Phi^k$. □

A common boost of all three states leaves the overlap unchanged, by the substitution $\nu \to \Lambda\nu$ in the integral. For two physical states in relative motion, the vertex still sources $\tau$-excited states with nonzero overlap in every case computed in Online Resource 1; this statement is computer-assisted and is not used elsewhere.

This is an instance of the known difficulties of constraints in interacting theories, from the loss of constraints to acausal propagation [53]; for bilocal fields the coupling must be arranged to respect the constraints, as has been done for a constant electric field [54]. A fallback keeps only the physical fields $\phi_n$ and couples them through the restriction $\Phi_{\text{phys}}(\sigma) = \Phi(\sigma, 0) = \sum_n \chi_n\,(0)\,\phi_n(\sigma)$ to the section. At $\nu = 0$ the boosted profile takes the rest-frame value $\chi_n(0) = \psi_n(0)\psi_0(0)^3$, independent of $\hat{p}$, so the coupling is local and Lorentz invariant, and only even levels couple, since $\psi_n(0) = 0$ for odd $n$. The Källén–Lehmann weights $\chi_{2k}(0)^2 = (\omega/\pi)^2\binom{2k}{k}4^{-k} \sim \omega^2\pi^{-5/2}k^{-1/2}$ decay while the levels are equally spaced in $M^2$, so the Euclidean two-point function of $\Phi_{\text{phys}}$ behaves as $\omega^{3/2}/\big(2\pi^{3/2}|p|\big)$ for $|p| \to \infty$. Thus $\Phi_{\text{phys}}$, of engineering dimension 3, has ultraviolet scaling dimension 3/2: the field $\widehat{\Phi} = (\pi/\omega)^{3/4}\Phi_{\text{phys}}$ has the two-point function $1/(2|p|)$ at large $|p|$. A cubic coupling $g\,\Phi_{\text{phys}}^3 = g(\omega/\pi)^{9/4}\widehat{\Phi}^3$ therefore has power-counting dimension $-1/2$, the superficial degree of divergence of a graph with $E$ external lines and $V$ vertices is $4 - 3/2\,E + 1/2\,V$, and the fallback is non-renormalizable by power counting (Online Resource 1, Sec. S2.8). Assessing this fallback as an effective theory requires a cutoff, a treatment of the infinite tower, and analyses of radiative corrections and vacuum stability.

### 7.3 Metric sector

**Proposition 25** (graviphotons and moduli of the normal block). *Consider the Kaluza–Klein reduction of the eight-dimensional Einstein–Hilbert action on $\Sigma \times N$ with constant fiber metric $h_N = -\eta$, without a compatibility condition relating the metric fluctuations to $K$, and with the overall sign of the action* *fixed by requiring a healthy four-dimensional graviton. In the sector independent of* $\nu$*, per unit volume of* $N$*:*

*(i)* *the graviphotons have kinetic term* $-1/4\,(h_N)_{ab}F^a\cdot F^b$*, so the three graviphotons along* $\tau_1,\tau_2,\tau_3$ *are ghosts;*

*(ii)* *the Einstein-frame kinetic form of the ten constant moduli* $\delta G_{ab}$*,*

$$\mathcal{K}(\delta G) = 1/4\ \mathrm{tr}(h_N^{-1}\delta G\, h_N^{-1}\delta G) + 1/8\,(\mathrm{tr}\, h_N^{-1}\delta G)^2, \tag{23}$$

*has signature* $(7,3)$*, and* $(6,3)$ *on the unimodular moduli; the negative directions are* $\delta G(\partial_{x_4},\partial_{\tau_i})$*,* $i=1,2,3$*.*

*Proof.* (i) is Proposition 8 with $\gamma = h_N$. (ii) By Appendix B, a modulus is healthy when $\mathcal{K}>0$ on it. In the basis $(\nu^0,\nu^1,\nu^2,\nu^3)$, in which $h_N=\mathrm{diag}(1,-1,-1,-1)$,

$$\mathcal{K}(\delta G) = 1/4\sum_a(\delta G_{aa})^2 + 1/8\,(\delta G_{00}-\sum_i \delta\, G_{ii})^2 - 1/2\sum_i(\delta G_{0i})^2 + 1/2\sum_{i<j}\left(\delta G_{ij}\right)^2.$$

The diagonal block is positive definite and the off-diagonal components separate, which gives $(7,3)$; the unimodular condition $\mathrm{tr}\,h_N^{-1}\delta G=0$ removes one positive diagonal direction. □

The signature $(7,3)$ is that of the symmetric space $\mathrm{GL}(4,\mathbb{R})/O(1,3)$, as for reductions on tori with a timelike circle, whose scalar cosets have indefinite signature [7]. The natural compatibility condition, $\delta h(K\,\cdot, K\,\cdot) = -\delta h$, does not remove these ghosts. It forces the normal block of the metric fluctuation to be minus the block along $\Sigma$, which locks the moduli to the four-dimensional metric, and the mixed block to be a 2-form, $B_{\mu\lambda} = \delta h\big(\partial_{\sigma^\mu},\partial_{\nu^\lambda}\big) = -B_{\lambda\mu}$ in the coordinates (9), in which $K\,\partial_{\sigma^\lambda} = \partial_{\nu^\lambda}$, so that the sixteen graviphoton components reduce to six. Imposed on the quadratic action, with the normalization and the $\nu$-independent truncation of Proposition 25, it gives $B$ the Lagrangian $-2L_{\mathrm{KR}} + 1/2\,(\partial\cdot B)^2$, which is the graviphoton kinetic term of Proposition 8 restricted to these six components; here $L_{\mathrm{KR}} = -1/12\,H_{\mu\nu\rho}H^{\mu\nu\rho}$ is the Kalb–Ramond Lagrangian, $H=dB$ and $(\partial\cdot B)_\nu = \partial^\mu B_{\mu\nu}$. No gauge symmetry can be used on these modes; the transverse polarization of $B$, the one that is physical in Kalb–Ramond theory, is a ghost; two further polarizations are dipoles, with propagator $\propto (p^2)^{-2}$ and solutions growing linearly in time; and the quadratic action of the symmetric block is not invariant under linearized four-dimensional diffeomorphisms. The statements above are computer-assisted, by computation in Online Resource 1, and are not used elsewhere. A dynamical $K$, or an action for a separate generalized metric of double-field-theory type, is outside the scope.

### 7.4 Scales

The flat structure $(h,K)$ is covariant under dilations, so it fixes neither $\omega$ nor $m_\Phi$. The tower (16) has equal spacing in $M^2$, unit multiplicity and spin zero, and only its even levels couple at $\nu=0$; these are structural properties of the free model and of the fallback coupling. Connecting the tower to observed particles would require a physical identification of the levels and an independently specified mass scale.

## 8 Relation to earlier proposals

The results above constrain the reading of Refs. [11–13] in which four-dimensional masses are eigenvalues associated with the temporal directions. With the full three-time Laplacian, the (3,3) Klein–Gordon equation $(\Delta_x - \Delta_t - m_\Phi^2)\Psi = 0$ gives $\Delta_t\Psi = -(|\mathbf{p}|^2 + m_\Phi^2)\Psi$ for a plane wave in $x$; the eigenvalue equation of Ref. [11], Eq. (14), reused in Ref. [12], Eq. (7), is therefore the bulk dispersion relation, and it does not define a mass generated by the temporal directions. In reductions satisfying Assumption 1, gradients along the extra times lower $M^2$ (Theorem 4); this applies equally to the temporal structure functions of Ref. [13], Eqs. (42)–(47). The para-Hermitian facts used here are elementary and proved in the text. The companion paper analyzes the four-dimensional sections of signature $(3,3)$; Theorem 4(b) generalizes its Eq. (A2) to warped fibers of arbitrary signature and every self-adjoint extension, and Corollary 5(iii) and Theorem 7 make clear the costs identified in its Proposition 27 and Remark 28 (closed timelike curves, or broken temporal translations).

# 9 Discussion

**Answer to the question.** For scalar mass operators satisfying Assumption 1, gradients along extra timelike directions contribute non-positively to the four-dimensional mass squared (Definition 2 and Theorem 4(a)). The mass spectrum on the unrestricted fiber Hilbert space is unbounded below (Theorem 4(b)); on a negative-definite fiber, levels above the warped bulk mass function require boundary terms (Theorem 4(c)); the harmonic confinement analyzed here produces an inverted tower with a bulk mass function that is tachyonic far from the origin (Corollary 5(iii)); a finite-multiplicity level under the homogeneity and boundary assumptions of Theorem 7 requires compact timelike directions; and the fiber-independent Einstein–Hilbert reduction has ghost graviphotons along the timelike directions (Proposition 8). Flat para-Hermitian space of signature $(3,3)$ carries no Lorentz algebra by automorphisms (Proposition 11). In signature $(4,4)$, where the Lorentz group is the stabilizer of a Lagrangian section, a local subsidiary condition yields a free positive tower. The positive part of this tower is carried by $x_4$, the $K$-partner of the time of the section, which is spacelike in the bulk metric, so that the mechanism is of Kaluza–Klein and bilocal type.

**Limitations.** The no-go statements are proved for free scalar fields with two-derivative actions on warped products. Theorem 4(b) holds for every self-adjoint extension; only part (c) assumes the absence of positive boundary terms. Spin-$1/2$ fields need a separate argument: squaring the Dirac operator reproduces the gradient sign, but we have not proved a theorem. Higher-derivative actions lie outside Assumption 1, and we have not analyzed them; in brane-world gravity with a Gauss–Bonnet term, timelike extra dimensions without tachyons or ghosts have been proposed [55]. Masses generated by interactions lie outside the scope of the identity, which constrains quadratic operators only; assessing an interacting completion requires a separate analysis of the constraints, the physical state space and stability. Every bulk considered here violates the chronology condition (Sec. 2.3). A different notion of mass, the ADM-type mass of initial data with several time dimensions, is the subject of a recent positive-mass theorem [56], which concerns the geometry of initial data.

**Outlook.** Two questions define the next steps toward an interacting theory based on the $(4,4)$ realization. The first is an interaction principle that preserves (13), for example through a gauge symmetry, a BRST construction or auxiliary fields. The second is a ghost-free metric sector. With fixed $K$, the linearized Einstein–Hilbert sector contains three ghost vectors and three ghost moduli (Proposition 25). The natural compatibility condition does not remove them: it reduces the graviphotons to a 2-form with a ghost polarization and dipole modes, and it locks the moduli to the four-dimensional metric, whose quadratic action then loses its linearized diffeomorphism invariance (Sec. 7.3, computer-assisted). A dynamical $K$, or a double-field-theory action for a separate generalized metric, remains open. Even if both questions were resolved, the resulting interacting theory would be a Born-geometric Kaluza–Klein model: its positive mass spacing would still be carried by a direction that is spacelike for the bulk kinetic metric, consistently with the sign identity. Both questions connect to Born reciprocity and metastring theory [18, 21, 52]. In the metaparticle the dual momentum enters the mass shell through the generalized metric, as $p^2 + \tilde{p}^2 + m^2 = 0$ (Ref. [52], Eqs. (5) and (17)), so that a spacelike $\tilde{p}$ raises the effective pole mass (Eq. (19) there) while a timelike $\tilde{p}$ enters with the opposite sign. Two-time physics [8, 10] is the best-developed framework in which a second time is made consistent, through an $\mathfrak{sp}(2,\mathbb{R})$ gauge symmetry, with exactly two times and with the mass carried by an extra spacelike direction; timelike T-duality relates theories with several times [6]. In all of these frameworks, as in the results above, the sign of a contribution to $M^2$ is the sign of the kinetic metric on the corresponding direction (the generalized metric for the metaparticle, $h$ here); which directions count as temporal depends on which metric is taken as kinetic.

## Supplementary information

Online Resource 1 (PDF) contains the derivations behind Remarks 6, 20 and 23 and the fallback two-point function of Sec. 7, the computations of the boosted overlaps after Proposition 24 and of the compatible metric sector of Sec. 7.3, and an index of Online Resource 2. Online Resource 2 is a zip archive (folder `nogo_supplement_scripts`) of computer-algebra and numerical scripts, with their outputs, that check the statements of Theorems 4 and 19, Corollaries 5 and 17, Remark 15, Lemmas 10 and 21, Propositions 8, 11, 12, 14, 16, 22, 24 and 25, and Appendixes A and B.

## Statements and declarations

**Funding.** The author received no funding for this work.

**Competing interests.** The author has no relevant financial or non-financial interests to disclose.

**Author contributions.** G. Kletetschka is the sole author and is responsible for the conceptualization, formal analysis, validation and writing (original draft, review and editing).

**Data availability statement.** All data generated or analyzed during this study are included in this article and its supplementary information files (Online Resources 1 and 2).

**Code availability statement.** The scripts used to check the results, with their complete outputs, a README file and a script that runs them all, are provided as Online Resource 2.

## Appendix A: Interaction overlaps

With $\psi_k(y) = (\omega/\pi)^{1/4}\left(2^k k!\right)^{-1/2} H_k\left(\sqrt{\omega}y\right)e^{-\omega y^2/2}$ and $z = \sqrt{\omega}\, y$,

$$I_k = \left(\frac{\omega}{\pi}\right)^{3/4} \frac{\omega^{-1/2}}{\sqrt{2^k k!}} \int_{\mathbb{R}} H_k\,(z)\, e^{-3z^2/2} dz.$$

The generating function $\sum_k H_k\,(z) t^k/k! = e^{2zt-t^2}$ and the Gaussian integral give

$$\sum_k \frac{t^k}{k!} \int_{\mathbb{R}} H_k\,(z) e^{-3z^2/2} dz = \sqrt{\frac{2\pi}{3}}\, e^{-t^2/3},$$

so the odd integrals vanish and $\int H_{2k}(z) e^{-3z^2/2} dz = \sqrt{2\pi/3}\,(2k)!\,(-1/3)^k/k!$. Hence $I_0 = \sqrt{2/3}\,(\omega/\pi)^{1/4} = \sqrt{6}\,\omega^{1/4}/\left(3\pi^{1/4}\right)$ and

$$\frac{I_{2k+2}}{I_{2k}} = -\frac{1}{3}\frac{(2k+2)(2k+1)}{k+1}\frac{1}{\sqrt{4(2k+2)(2k+1)}} = -\frac{1}{3}\sqrt{\frac{2k+1}{2k+2}},$$

which is (22). Numerically $I_0, I_2, I_4, I_6 = 0.613, -0.145, 0.042, -0.013$ in units of $\omega^{1/4}$. For a monomial vertex $\Phi^k$ the corresponding overlap with two $\tau$ quanta is

$$\int_{\mathbb{R}} \psi_0\,(y)^{k-1}\psi_2(y)\,dy = \left(\frac{\omega}{\pi}\right)^{k/4} \frac{\omega^{-1/2}}{\sqrt{8}} \int_{\mathbb{R}} (4z^2-2)\; e^{-kz^2/2} dz = \left(\frac{\omega}{\pi}\right)^{k/4} \frac{\omega^{-1/2}}{\sqrt{2}} \sqrt{\frac{2\pi}{k}}\;\frac{2-k}{k},$$

which is nonzero for every $k \geq 3$.

## Appendix B: Kaluza–Klein formulas

Consider metrics on $\mathbb{R}^{3,1} \times F$ that are independent of the fiber coordinates $y$,

$$ds^2 = g_{\mu\nu}(x)dx^\mu dx^\nu + G_{ab}(x)\big(dy^a + A^a_\mu dx^\mu\big)\big(dy^b + A^b_\nu dx^\nu\big). \tag{B1}$$

For non-compact $F$ this is a consistent truncation and the action is taken per unit fiber volume; on a torus it is the zero-mode sector. Write $\psi_\mu = G^{-1}\,\partial_\mu G$ (a matrix) and $V = \sqrt{|\det G|}$. For $A = 0$ the scalar curvature is

$$R = R_g - g^{\mu\nu}\big[\mathrm{tr}\big(G^{-1}\nabla_\mu\nabla_\nu G\big) - 3/4\ \mathrm{tr}\big(\psi_\mu\psi_\nu\big) + 1/4\ \mathrm{tr}\,\psi_\mu\,\mathrm{tr}\,\psi_\nu\big], \tag{B2}$$

which for a one-dimensional fiber $G = e^{2f}$ reduces to the warped-product formula $R = R_g - 2(\Box f + (\partial f)^2)$. Using $\mathrm{tr}(G^{-1}\Box G) = \nabla_\mu\,\mathrm{tr}\,\psi^\mu + \mathrm{tr}\big(\psi_\mu\psi^\mu\big)$ and $\partial_\mu V = 1/2\,V\,\mathrm{tr}\,\psi_\mu$, and integrating by parts,

$$V\sqrt{-g}\,R \simeq V\sqrt{-g}\big[R_g - 1/4\ \mathrm{tr}\big(\psi_\mu\psi^\mu\big) + 1/4\ \mathrm{tr}\,\psi_\mu\,\mathrm{tr}\,\psi^\mu\big] - 1/4\,V\sqrt{-g}\,G_{ab}F^a_{\mu\nu}F^{b\,\mu\nu}, \tag{B3}$$

where the last term is the standard graviphoton contribution [40], with $F^a = dA^a$. The overall sign of the action is fixed by requiring a healthy four-dimensional graviton, so that the action is a positive multiple of $\int V\sqrt{-g}\,R$. For constant $G = \gamma$ this gives the kinetic term $-1/4\,\gamma_{ab}F^a_{\mu\nu}F^{b\,\mu\nu}$ of Proposition 8. In four dimensions the Weyl rescaling $g_{\mu\nu} = V^{-1}g^E_{\mu\nu}$ removes the factor $V$ in front of $R_g$ and produces $-6\big(\partial\ln V^{1/2}\big)^2 = -3/8\,(\mathrm{tr}\,\psi)^2$, while the moduli terms in (B3) are invariant. The Einstein-frame Lagrangian is

$$\mathcal{L}_E = \sqrt{-g_E}\big[R_E - 1/4\ \mathrm{tr}\big(\psi_\mu\psi^\mu\big) - 1/8\ \mathrm{tr}\,\psi_\mu\,\mathrm{tr}\,\psi^\mu\big] + \cdots, \tag{B4}$$

so the kinetic form of the moduli $\delta G$ at a background $\gamma$ is

$$\mathcal{K}(\delta G) = 1/4\ \mathrm{tr}(\gamma^{-1}\delta G\,\gamma^{-1}\delta G) + 1/8\,(\mathrm{tr}\,\gamma^{-1}\delta G)^2, \tag{B5}$$

and a modulus is healthy when $\mathcal{K} > 0$ on it. For positive-definite $\gamma$, $\mathcal{K}$ is positive definite, as it must be for a torus. For a one-dimensional fiber, $G = e^{2f}$, (B4) reduces to $R_E - 3/2\,(\partial f)^2$, the familiar radion kinetic term.